\pdfoutput=1 % For pdflatex
\documentclass[english,11pt,a4paper]{article}

\usepackage{jheppub} 
\usepackage{graphicx}% Include figure file
\usepackage{bm}% bold math
\usepackage{amsmath}
\usepackage{lineno}
\usepackage{amssymb}
\usepackage[utf8]{inputenc}
\usepackage{lipsum}

\IfFileExists{dsfont.sty}
	{\usepackage{dsfont}
         \let\mathbb=\mathds
         \newcommand{\id}{\mathds{1}}}
	{\typeout{Package dsfont.sty was not found, using alternative macros.}
         \let\mathds=\mathbb
         \newcommand{\id}{\mbox{1 \kern-.59em {\rm l}}}}

\renewcommand\a{\alpha}
\renewcommand\b{\beta}
\renewcommand\d{\delta}

\renewcommand\l{\lambda}

\renewcommand\t{\tau}

\renewcommand\o{\omega}

\newcommand\e{\epsilon}
\newcommand\g{\gamma}

\newcommand\m{\mu}
\newcommand\n{\nu}

\newcommand\p{\pi}
\newcommand\h{\eta}
\newcommand\s{\sigma}

\newcommand\w{\omega}
\renewcommand\P{\Pi}

\renewcommand{\part}{{\rm part}}
\newcommand{\be}{\begin{equation}}
\newcommand{\ee}{\end{equation}}
\newcommand{\bes}{\begin{subequations}}
\newcommand{\ees}{\end{subequations}}
\newcommand{\bea}{\begin{eqnarray}}
\newcommand{\eea}{\end{eqnarray}}

\newcommand{\pa}{\partial}

\newcommand{\na}{\nabla}
\def\nbox#1#2{\vcenter{\hrule \hbox{\vrule height#2in
\kern#1in \vrule} \hrule}}
\def\sq{\,\raise.5pt\hbox{$\nbox{.10}{.10}$}\,}
\def\sqb{\,\raise.5pt\hbox{$\overline{\nbox{.09}{.09}}$}\,}
\allowdisplaybreaks
\begin{document}

\preprint{LA-UR-20-20237}

\title{Drag force to all orders in gradients}

\author[a,b,c]{Jared Reiten}
\author[c]{and Andrey V. Sadofyev}

\affiliation[a]{Department of Physics and Astronomy, University of California, 475 Portola Plaza, Los Angeles, CA 90095, U.S.A.}
\affiliation[b]{Mani L. Bhaumik Institute for Theoretical Physics, University of California, 475 Portola Plaza, Los Angeles, CA 90095, U.S.A.}
\affiliation[c]{Theoretical Division, MS B283, Los Alamos National Laboratory, Los Alamos, NM 87545, U.S.A.}

\emailAdd{jdreiten@physics.ucla.edu}
\emailAdd{sadofyev@lanl.gov}

\abstract{
We study the energy loss of a heavy quark slowly moving through an evolving strongly coupled plasma. We use the linearized fluid/gravity correspondence to describe small perturbations of the medium flow with general spacetime dependence. This all order linearized hydrodynamics results in a drag force exerted on a heavy quark even when it is at rest with the fluid element. We show how the general contribution to the drag force can be derived order by order in the medium velocity gradients and provide explicit results valid up to the third order. We then obtain an approximate semi-analytic result for the drag force to all orders in the gradient expansion but linearized in the medium velocity. Thus, the effects of a class of hydrodynamic gradients on the drag force are re-summed, giving further insight into the dissipative properties of strongly coupled plasmas. The all order result allows us to study the drag force in the non-hydrodynamic regime of linear medium perturbations that vary rapidly in space and time. 
}

\date{\today}
\maketitle

\section{Introduction}
The strong energy loss of partons propagating through the quark-gluon plasma (QGP) produced in heavy-ion collisions (HIC) is a distinctive feature of heavy-ion experiments. This phenomenon, known as jet quenching, has attracted significant attention in the literature -- for a review see \cite{Busza:2018rrf}. If one treats the QGP as weakly coupled, then parton energy loss can be successfully described with perturbative techniques based on fundamental QCD \cite{Dokshitzer:2001zm, Baier:2001yt, Jeon:2003gi, Mustafa:2003vh, Djordjevic:2003qk, Djordjevic:2003zk, Zhang:2003wk, Moore:2004tg, Wicks:2005gt, Vitev:2007ve, Kang:2016ofv, Blagojevic:2018nve}. However, the plasma produced in HIC exhibits strong collective phenomena \cite{Adcox:2004mh, Arsene:2004fa, Back:2004je, Adams:2005dq, Aamodt:2010pa, ATLAS:2011ah, Chatrchyan:2012ta} and, at later stages, is described by relativistic hydrodynamics -- for a review see \cite{Romatschke:2017ejr} -- with parameters indicating its strongly coupled nature. 

While there is no reliable tool to describe QCD plasma at all length scales, insights into how strongly coupled dynamics affect parton energy loss can be accessed through the use of holographic duality. This technique allows one to relate calculations in non-perturbative QFT to their analogues in a dual higher-dimensional gravitational theory. The dual description of QCD is still unknown and therefore one has to utilize some QCD-like theory with a known holographic dual as a model. The simplest example of such a model is given by strongly coupled $\mathcal{N}=4$ supersymmetric Yang-Mills (SYM) in the large-$N_c$ limit which can be described under AdS/CFT correspondence \cite{Maldacena:1997re, Gubser_1998, Witten:1998qj}. While $\mathcal{N}=4$ SYM differs considerably from QCD, its plasma phase shares similarities with that of QCD in the strongly coupled regime. This plasma is dual to classical gravity in a $4+1$-dimensional asymptotically anti-de-Sitter (AdS) spacetime with a black brane horizon needed to introduce a finite temperature scale, see e.g. \cite{CasalderreySolana:2011us, DeWolfe:2013cua}. Such a holographic approach, and its extensions, have been used to study the energy loss of a heavy quark propagating through the holographic plasma in a variety of setups, as well as to study the behavior of other probes of the plasma, see e.g. \cite{Herzog:2006gh, Gubser:2006bz, CasalderreySolana:2006rq, Gubser:2006nz, Liu:2006he, Liu:2006ug, Liu:2006nn, Caceres:2006ta, Gubser:2008as, Chesler:2008uy, Hatta:2008tx, Fadafan:2008bq, Arnold:2010ir, Chesler:2011nc, Arnold:2012qg, Arnold:2012uc, Chernicoff:2012bu, Giataganas:2012zy, Chesler:2013urd, Lekaveckas:2013lha, Ficnar:2013qxa, Dudal:2014jfa, Sadofyev:2015hxa, Rajagopal:2015roa, Chesler:2015nqz, Mamo:2016xco, Finazzo:2016mhm, Dudal:2016joz, Li:2016bbh, Singh:2017nfa, Dudal:2018rki, Bohra:2019ebj} and references therein. Most recently, these studies of medium-probe interactions evolved into several models of jets propagating through a strongly coupled QGP which incorporate some aspects of both weakly and strongly coupled dynamics \cite{Casalderrey-Solana:2014bpa, Casalderrey-Solana:2015vaa, Rajagopal:2016uip, Casalderrey-Solana:2016jvj, Hulcher:2017cpt, Brewer:2017dwd, Brewer:2017fqy, Brewer:2018mpk, Casalderrey-Solana:2019ubu}.

However, the QGP produced in HIC undergoes a highly non-trivial evolution starting as far-from-equilibrium matter immediately after the collision, then ultimately transitioning to a thermalized system following hydrodynamic equations. In \cite{Lekaveckas:2013lha, Rajagopal:2015roa} it was shown that the energy loss of a heavy quark is modified in a flowing, non-static plasma and that the resulting drag force depends on hydrodynamic gradients. These contributions influence the behavior of the energy loss in moving matter and are important for the phenomenology of HIC, as measuring the energy loss of heavy quarks and other probes can in principle allow one to access details of the medium motion during its evolution. On the other hand, by studying the dependence of the drag force on gradients in the medium, one can additionally access details of the dissipative properties of strongly interacting plasmas. 

The hydrodynamic evolution of the holographic plasma can be described with the gravitational dual obtained order by order in the gradient expansion \cite{Bhattacharyya:2008jc, Erdmenger:2008rm, Banerjee:2008th}. The dual description of a heavy quark is given by a classical string whose end point is attached to the AdS boundary and moving, in the simplest case, at a constant speed \cite{Herzog:2006gh, Gubser:2006bz, CasalderreySolana:2006rq}. The string profile is sensitive to the bulk metric and so, too, is the drag force. In \cite{Lekaveckas:2013lha, Rajagopal:2015roa} the leading corrections to the drag force in neutral and charged holographic plasmas, respectively, were obtained by studying the string in the background metric of a flowing holographic plasma to the first order in gradients. In \cite{Rajagopal:2015roa} the drag force acting on a quark at rest in the fluid element rest frame was used as a probe of dissipation sourced by anomalous transport \cite{Kharzeev:2015znc}. 

The gradient expansion of the hydrodynamic dual has two types of contributions -- terms that are linear in the medium velocity with multiple gradients involved (for example $\na\na\cdots\na u$), and nonlinear terms (such as $(\na u)^2$). Recently, it was suggested that the former terms can be taken into account to all orders in the gradient expansion within the framework of the linearized fluid/gravity correspondence \cite{Bu:2014sia, Bu:2014ena, Bu:2015ika, Bu:2015bwa}, see also \cite{Lublinsky:2007mm, Lublinsky:2009kv}. 

In this work, we continue the study of the drag force exerted on a heavy quark propagating through a holographic plasma, restricting our consideration to small quark velocities. In particular, we focus on the higher order gradient corrections to the drag force from the linearized hydrodynamics of the strongly coupled plasma, which provide a benchmark for higher-order hydrodynamic effects on other probes. Such corrections are, for instance, important for the relativistic hydrodynamic description of the QGP in HIC, which usually involves terms at least up to the second order to ensure causality of the theory \cite{Romatschke:2017ejr}. We utilize the linearized fluid/gravity correspondence to derive the equations governing the string profile in the corresponding background bulk metric, assuming the perturbations due to the medium motion to be small. We compare these equations in the linearized background to the ones obtained to first order in gradients and show how they can be used to analytically derive the drag force order-by-order in gradients linear in the medium velocity. Here the first three orders in this expansion are given explicitly. We also consider how a non-trivial quark trajectory affects the drag force to linear order in the amplitude of the quark displacement. We further use the matching procedure to stitch the series expansions in the holographic coordinate at the boundary and horizon, deriving an approximate solution for the drag force in the linearized background to all orders in gradients. Finally, we analyze the behavior of the drag force as a function of the medium velocity gradients, compare the approximate but all order result with the leading terms in the gradient expansion, and comment on the features of gradient effects, such as the fluid element rest frame choice. While linearized hydrodynamics supports only a set of modes with fixed dispersion $\o_n(k)$, where $n$ enumerates the modes, we collect the general information on the drag force dependence on $\o$ and $k$ separately. This allows one to additionally take into account the case of evolution under external forces. In this way, we provide further insights into heavy quark energy loss within evolving strongly coupled plasmas.

This work is organized as follows. In Section \ref{sec:Fluid} we briefly review and compare the holographic dual of a slowly varying fluid up to first order in gradients and that of linearized all order hydrodynamics. In Section \ref{sec:DragOld} we set up the drag force calculation within the AdS/CFT framework and show how the linearized hydrodynamic dual can be used to derive (linear) higher order gradient corrections along the logic of \cite{Lekaveckas:2013lha, Rajagopal:2015roa}. In Section \ref{sec:DragNew} we proceed by solving for the all order drag force as functions of the medium velocity Fourier amplitude, as well as show how this result is related to the drag force on a heavy quark undergoing some arbitrary motion of small amplitude. Finally, in Section \ref{sec:Summary} we review our results and discuss their physical interpretation.

\section{Fluid/gravity correspondence}
\label{sec:Fluid}
In this section we review the fluid/gravity duality \cite{Bhattacharyya:2008jc, Erdmenger:2008rm, Banerjee:2008th} and its linearized all order extension \cite{Bu:2014sia, Bu:2014ena} for $\mathcal{N}=4$ SYM in order to derive all the elements required for the drag force calculation. The dual theory is described by the 4+1-dimensional Einstein action
\begin{eqnarray}
S=-\frac{1}{16\pi G_5}\int d^5 x \sqrt{-G}\left(R+12\right)\,,
\end{eqnarray}
where $G_5$ is the 4+1-dimensional Newton constant and we have set $R_{AdS}=1$. The corresponding equations of motion read
\begin{eqnarray}
\label{EinsteinEqn}
R_{MN}+4G_{MN}=0
\label{eq:EofM}
\end{eqnarray}
and have a static uniform black brane solution
\begin{eqnarray}
\label{g0}
ds^2&=&-r^2f(r)u_\mu u_\nu dx^\mu dx^\nu+r^2P_{\mu\nu}dx^\mu dx^\nu-2u_\mu dx^\mu dr\,,
\end{eqnarray}
where $f=1-\frac{\p^4 T^4}{r^4}$ defines the horizon at $r_h=\p T$, $u^\m$ is a constant boost vector satisfying $u^2=-1$, and $P_{\m\n}$ is the projector onto directions transverse to the fluid 4-velocity, $P_{\m\n}=\eta_{\m\n}+u_\m u_\n$. In the boundary theory, this solution corresponds to a static plasma at constant temperature $T$. In order to simplify the form of all resulting equations, we set $\p T=1$ and hence work in units of $\p T$. Geometrically, this sets the location of the black brane horizon to $r=\p T=1$ in our coordinate system. Here $G_{MN}$ is the bulk metric defining  the line element $ds^2$. We use capital Latin indices for the 4+1-dimensional bulk coordinates, lowercase Greek indices for the boundary 3+1-dimensional space, lowercase Latin indices $i,j,k,...$ for the spatial coordinates of the boundary, and lowercase Latin indices $a,b,c,...$ for the worldsheet coordinates of the string corresponding to a heavy quark in the dual description.

To find the metric dual to a flowing strongly coupled plasma, one has to allow the temperature and velocity to be functions of the boundary coordinates. However, obtaining an exact solution to such a problem is challenging, so it is natural to rely on some expansion about the known static/uniform solution. The choice of such an expansion is the main difference between the usual hydrodynamics based on the gradient expansion and all order linearized hydrodynamics. In standard hydrodynamics, the expansion is performed in smallness of the gradients -- $T(x^\a)$ and $u_\m(x^\a)$ are slowly varying functions and can thus be safely expanded about a given point in the boundary coordinates. In linearized hydrodynamics, the smallness parameter is the amplitude of the $3$-dimensional velocity, which also fixes the smallness of variations in $T$. The gradients are then assumed to be of the same order. 

\subsection{First order hydrodynamics}
If the gradient expansion is utilized, then the dual of the flowing fluid can be found in a fashion analogous to the way that hydrodynamics is typically derived itself. One starts by allowing the hydrodynamic variables in the static and uniform solution to slowly vary from point to point in the boundary coordinates. Upon doing so, the modified metric no longer provides a solution to the Einstein equation and thus needs to be corrected. The required correction may then be determined order by order in smallness of the gradients
\begin{equation}
G_{MN}=G^{(0)}_{MN}+G^{(1)}_{MN}+{\cal O}(\partial^2)\,,
\end{equation}
where $G^{(0)}_{MN}$ is defined in (\ref{g0}) and $G_{MN}^{(1)}$ contains all possible gradient structures made of $u^\m$ and $T$ to first order. The solution up to first order was obtained in \cite{Bhattacharyya:2008jc}, which in the ingoing Eddington-Finkelstein coordinates reads
\begin{eqnarray}
\label{metric}
ds^2&=&-r^2f(r)u_\mu u_\nu dx^\mu dx^\nu+r^2P_{\mu\nu}dx^\mu dx^\nu-2u_\mu dx^\mu dr+r^2F(r)\sigma_{\mu\nu}dx^\mu dx^\nu\notag\\&~&\qquad+r^2j_\sigma\left(P^\sigma_\mu u_\nu+P^\sigma_\nu u_\mu\right)dx^\mu dx^\nu+\frac{2}{3}r(\partial\cdot u) u_\mu u_\nu dx^\mu dx^\nu
\end{eqnarray}
where
\begin{equation}
\sigma^{\mu\nu}\equiv  P^{\mu\alpha}P^{\nu\beta} \left( \partial_\alpha u_\beta +\partial_\beta u_\alpha \right)
-\frac{2}{3}P^{\mu\nu}\partial\cdot u~~~,~~~j_\sigma\equiv -\frac{1}{r}(u\cdot\partial) u_\sigma\,,
\end{equation}
and the function $F(r)$ is given by
\begin{eqnarray}\label{Fdefn}
F(r)&\equiv&\frac{1}{4}\left[2\tan^{-1}\left(\frac{1}{r}\right)-\log\left(\frac{r^4}{(1+r)^2(1+r^2)}\right)\right]\,.
\end{eqnarray}
Note that $f(r)$ implicitly depends on $x^\m$ through temperature variations. The solution of the Einstein equation is not unique due to residual freedom -- freedom which can be eliminated by fixing the definition of $4$-velocity in the dual hydrodynamics at each order in the gradient expansion. Here the dissipative (first order) contributions to the fluid stress energy tensor are required to be transverse to $u^\m$. Thus, the local rest frame defined with $u^\m$ is the rest frame of the energy density (fluid momentum vanishes), known as the Landau frame \cite{Landau:1986aog}. The Landau frame, which can be straightforwardly generalized in all order linearized hydrodynamics, is used throughout this paper. 

Utilizing the holographic dictionary, one may extract the kinetic coefficients governing the hydrodynamic evolution of the strongly coupled plasma. Particularly, the shear viscosity is proportional to the term of $F(r)$ in the near-boundary expansions scaling as $r^{-4}$ and, since $F(r)\simeq \frac{1}{r}-\frac{1}{r^4}\frac{\eta_0}{4q}$, in our dimensionless notations $F(r)\simeq \frac{1}{r}-\frac{1}{4r^4}$ with $\eta_0=q(\p T)^3=q$, where the subscript ``$0$'' is used to distinguish from the all order $\eta\left(\pa_t,\pa_i^2\right)$ and the factor $q$ is defined by the rank of the gauge group. The equation of state of the holographic plasma reads $\e=3P=3q(\p T)^4=3q$ resulting in $s=\frac{\pa P}{\pa T}=4\p q (\p T)^3=4\p q$. Thus, one can easily reproduce the famous relation
$$\frac{\eta_0}{s}=\frac{1}{4\p}\, ,$$
giving the renowned property of strongly coupled plasmas \cite{Policastro_2001,Policastro_2002, Kovtun_2005}.

In \cite{Lekaveckas:2013lha, Rajagopal:2015roa} the string profile was studied in this background metric (\ref{metric}) and solved to the same order of accuracy. Thus, the drag force was derived to the first order in the gradient expansion, incorporating the corresponding effects of the medium motion. Curiously, it was shown that for a fast quark the gradient effects can become large, breaking the gradient expansion.

\subsection{All order linearized hydrodynamics}
Turning to linearized all order hydrodynamics, one promotes the hydrodynamic variables in the static and uniform solution to arbitrary functions of the boundary coordinates, while maintaining the requirement that these perturbations are small in amplitude. In this case, it is more natural to introduce the $3$-dimensional fluid velocity $\b_i$ since $u_\m\simeq\left(-1, \b_i\right)+\mathcal{O}(\b^2)$ and expand
\bea
G_{MN}=G^{(0)}_{MN}+G^{(1)}_{MN}+\mathcal{O}(\b^2)\,,
\label{G01lin}
\eea
with the temperature fluctuation being of the same order in smallness as $\b_i$. The fluid/gravity solution of this linearized problem was obtained in \cite{Bu:2014sia, Bu:2014ena}, which in the ingoing Eddington-Finkelstein coordinates reads
\bea
\label{metricallorder}
ds^2&=&-r^2f(r)u_\mu u_\nu dx^\mu dx^\nu+r^2P_{\mu\nu}dx^\mu dx^\nu-2u_\mu dx^\mu dr\notag\\&~&\qquad+\frac{K}{r^2}dt^2+2j_idtdx^i+r^2\a_{ij} dx^i dx^j\,,
\eea
where $K$ satisfies
\bea
\label{keq}
3r^2\pa_r K=6r^4\pa\cdot \b+r^3\pa_t\pa\cdot\b-2\pa\cdot j-r\pa_r\pa\cdot j-r^3\pa_i\pa_j\a_{ij}\,.
\eea
The vector and tensor sectors of the metric can be uniquely parametrized with
\bea
\label{vectorsector}
j_i&=&a(\pa_t, \pa_i^2, r)\b_i+b(\pa_t,  \pa_i^2, r)\pa_i\,\pa\cdot\b\notag\\
\a_{ij}&=&c(\pa_t, \pa_i^2, r)\s_{ij}+d(\pa_t,  \pa_i^2, r)\p_{ij}\,,
\eea
where $\s_{ij}$ is the spatial part of $\s_{\m\nu}$ defined above and $\p_{ij}=\left(\pa_i\pa_j-\frac{1}{3}\d_{ij}\pa^2\right)\pa\cdot\b$. The decomposition in (\ref{vectorsector}) then reduces the dynamical Einstein equations into a set of four coupled second-order ODEs in the holographic coordinate $r$ for the four Fourier transformed coefficient functions $(a,b,c,d)$, while $K$ can be obtained from the decoupled equation (\ref{keq}) if the other coefficient functions are known, see \cite{Bu:2014ena}. 

In this all order hydrodynamics, it is convenient to introduce two viscosities which are now functions of the gradients or, in the Fourier transformed descriptions, functions of $\o$ and $\vec k$. As in first order hydrodynamics, the viscosities are given by the appropriate terms \cite{Bu:2014sia, Bu:2014ena} in the near-boundary expansion\footnote{Note that our normalization for $\h (\w,k^2)$ differs from that of the authors of \cite{Bu:2014ena} by a factor of 1/2, resulting in $\h(\w,k^2)\rightarrow\h _0\equiv1$ as $\w,k^2\rightarrow0$, to match the notations of \cite{Bhattacharyya:2008jc}.}
\bea
c(r)=\frac{1}{r}-\frac{\eta(\o, k^2)}{4 r^4}+\mathcal{O}\left(\frac{1}{r^5}\right)~~~,~~~d(r)=-\frac{\zeta(\o, k^2)}{4 r^4}+\mathcal{O}\left(\frac{1}{r^5}\right)\label{cdbdy}\,.
\eea
For small frequency and momentum, the shear viscosity agrees with $\eta_0$ of first order hydrodynamics, while $\zeta$ appears only at $\mathcal{O}\left(\pa^3\right)$ of the $T^{\m\nu}$ expansion. In the opposite limit of large frequencies, the two viscosities go to zero as negative powers of frequency.

In our computation of the drag force on a heavy quark, we will rely on this expansion and adopt the metric (\ref{metricallorder}) for this purpose. In solving the string equations of motion, we expand the string profile in the same manner, treating the perturbations due to the medium motion to be small in amplitude.

\section{Drag force: static medium}
\label{sec:DragOld}
Turning now to the calculation of the drag force exerted on a heavy impurity in a strongly coupled fluid, we consider the dual of a quark given by the end point of a string which hangs down to the horizon \cite{Herzog:2006gh, Gubser:2006bz}. In the infinite mass limit, the end point of the string is attached to the AdS boundary. The dynamics of the classical string are described by Nambu-Goto action
\begin{eqnarray}
S_{NG}=-\frac{\sqrt\lambda}{2\pi}\int d\tau d\sigma\sqrt{-g(\tau,\sigma)}\,,
\end{eqnarray}
where $\l=g^2 N_c$ is the `t Hooft coupling defining the string tension and is assumed to be large, $g(\tau,\sigma)=\det g_{ab}(\tau,\sigma)$ is the determinant of the induced world-sheet metric $g_{ab}=G_{AB}\pa_a X^A(\tau,\sigma)\pa_b X^B(\tau,\sigma)$ with $G_{AB}$ being function of $X^M$, and $\s^a=(\t, \s)$ gives the worldsheet coordinates. For further convenience, we use the freedom in string parametrization to choose $t(\t,\s)=\t$ and $r(\t,\s)=\s$. The resulting string equations of motion then read
\begin{eqnarray}
\partial_\tau\left(\frac{\delta {\cal L}}{\delta\partial_\tau \vec X}\right)+\partial_\sigma\left(\frac{\delta {\cal L}}{\delta\partial_\sigma \vec X}\right)=\left(\frac{\delta {\cal L}}{\delta \vec X}\right)\ .
\label{eq:EulerLagrange}
\end{eqnarray}
Given a particular motion of the string end point at the boundary, once the solution for the string profile is obtained the drag force is calculated through
\bea
f^\m(\t)\equiv-\lim_{\s\to\infty}\,\eta^{\m\n}\P^\s_\n(\t,\s)\,,
\label{eq:dragforcedef}
\eea
where $\P^\s_M\equiv\frac{\d \cal L}{\d\pa_\s X^M }$ is one of the canonical string momenta fluxes and the force is defined at the given location along the heavy quark trajectory. One should note here that the force is not a Lorentz $4$-vector by definition, but its transformation properties can be easily obtained \cite{Lekaveckas:2013lha}.

For a heavy quark at rest in the rest frame of the static uniform plasma, described by $G^{(0)}_{MN}$ of (\ref{G01lin}), the string profile is trivial with $\vec{x}(\t,\s)=0$. If the quark is moving with a constant velocity for a long time, the string is expected to take the so-called ``trailing" shape \cite{Herzog:2006gh, Gubser:2006bz} given by
\begin{equation}
\vec x(\tau,\sigma)=\vec v\,\tau+\vec \xi(\sigma)\,,
\end{equation}
where $\vec v$ is the velocity of the heavy quark. Analyzing the string equations of motion, one may note that the momentum conjugate to $\vec \xi$ is constant -- the solution reads
\bea
\vec\xi(\s)=-{\vec v}\left[\tan^{-1}(\s)-\frac{\p}{2}\right]\,.
\eea
The integration constants are fixed by requiring that the string end point moves along trajectory $\vec x=\vec v t$ at the boundary and that the string profile is regular at the worldsheet horizon $\s= r_h \sqrt{\g}= \sqrt{\g}$ with $\g=(1-v^2)^{-1/2}$.
 
The drag force exerted on the heavy quark moving through the static uniform holographic plasma is readily obtained from (\ref{eq:dragforcedef}) and is given by
\begin{eqnarray}
\vec{f}^{(0)}=\frac{\sqrt{\lambda}}{2\p} \g\,\vec v\,.
\label{OrdinaryDrag}
\end{eqnarray}
In what follows, we will treat the quark velocity to be small -- linearizing our consideration in this parameter.

\section{Drag force: fluid motion}
\label{sec:DragNew}
In this section we derive the drag force felt by a heavy quark moving through the holographic plasma, as described by the all order linearized fluid/gravity correspondence. We consider the case of a small quark velocity, then, as we will see, the drag force has two separate contributions due to the motion of the quark and that of the medium. In this regime, the linearized equations governing the string profile can be used to derive the gradient corrections order by order. We then give a brief description of the procedure introduced in \cite{Gregory:2009fj} that is based on matching the boundary and horizon series solutions of the holographic equations at some intermediate point. With this tool, we derive approximate solutions for the $(a,b,c,d)$ functions, following \cite{Bu:2014ena}, and for the string profile in this background. Then the general drag force can be straightforwardly obtained in Fourier space as a function of frequency and linear momentum. Again, note that we work in units where $\p T=1$, and hence all instances of $\o$, $\vec k$, and $\vec f$ will be dimensionless -- physical values will be understood to be expressed in units of the appropriate powers of $\p T$.

\subsection{Gradient expansion}
In the limit of small quark velocity, the string profile can be decomposed as
\bea
\vec x(\t,\s)=\vec{x}^{(0)}(\t,\s)+\vec{x}^{(1)}(\t,\s)~~~,~~~\vec{x}^{(1)}(\t,\s)=\vec{x}^{(1)}_{b}(\t,\s)+\vec{x}^{(1)}_{s}(\t,\s)\,,
\eea
where the second term in $\vec x(\t,\s)$ is considered to be small. Without loss of generality, we take the zeroth order solution to be a straight string hanging down from the center of the boundary coordinate system to the horizon, $\vec{x}^{(0)}(\t,\s)=0$, which satisfies the string equations of motion in the background metric $G^{(0)}_{MN}$. For further convenience, we also split the leading correction to the string profile into the homogeneous part $\vec{x}^{(1)}_{b}(\t,\s)$, which solves the homogeneous equations with the boundary condition $\vec{x}^{(1)}_{b}(\t,\s)|_{\s\to\infty}=\vec{y}(\t)$, and the solution of the non-homogeneous equations $\vec{x}^{(1)}_{s}(\t,\s)$, which cancels the contribution of $G^{(1)}_{MN}$ into the equations of motion and satisfies $\vec{x}^{(1)}_{s}(\t,\s)|_{\s\to\infty}=0$. 

In this linearized setup, it is convenient to Fourier transform the metric perturbation as well as the string profile
$$ G^{(1)}_{MN} = \int\frac{d^4 k}{(2\p)^4} G^{(1)}_{MN}(k_\m, \s) e^{i k_\m x^\m}~~~,~~~\vec x= \int\frac{d\o}{2\p}\vec x(\o, \s) e^{-i \o \t}\,,$$
where $k^\m=(\o, \vec k)$. To maintain notational brevity throughout the text, we will use the same symbols for coordinate and momentum space functions. This results in the linearized string equations of motion taking the form
\bea
\mathcal{D}_{\s}\vec x^{(1)}=\vec\b\,(i\o-2\s)+\vec\b\, \pa_\s \frac{a}{\s^2}-\vec k \left(k\cdot\b\right)\,\pa_\s\frac{b}{\s^2}\,,
\label{stringeoms}
\eea
where the source on the r.h.s. depends on the form of the metric (\ref{metricallorder}) and all functions should be understood as the Fourier amplitudes depending on $r$, $\o$ and $\vec k$. In the above equation, $\mathcal{D}_{\s}$ is the differential operator defined by
\bea
\mathcal{D}_{\s}\equiv(1-\s ^4)\pa _{\s}^2+(2 i \o \s ^2-4\s ^3)\pa _{\s}+2 i \o \s \, .
\eea
Note that the drag force is sensitive only to $j_i$ components of the metric perturbation in the limit of small quark velocity, as manifested by the appearance of only the $a$ and $b$ coefficient functions in (\ref{stringeoms}). This is simply due to the fact that there is no other 3-vector in the bulk metric to first order in perturbations.  

Let us take a closer look at the corrections to the drag force due to motion of the medium and constrain the quark to be at rest. The string profile equations (\ref{stringeoms}) provide a convenient framework for obtaining gradient corrections to the string profile that corresponds to an \textit{additional} expansion in smallness of $\o$ and $\vec k$ -- this allows us to make contact with previous studies in the literature and also extend them by including the higher order gradients (for terms linear in the medium velocity). At zeroth order in gradients, the string equations of motion are identical to the case of the trailing string with the quark velocity replaced by the minus medium velocity, c.f. \cite{Lekaveckas:2013lha}. The string profile reads
\bea
\vec{x}_s^{(1,0)}=\vec\b\int_{\infty}^\s  \frac{d\s'}{1+\s'^2}=\vec\b\left[\tan^{-1}(\s)-\frac{\p}{2}\right]\,,
\eea
where the superscript $(1,m)$ denotes the contribution that is linear in perturbation amplitude and at $m^{\rm th}$ order in gradients. One integration constant is fixed by regularity at the worldsheet horizon, which coincides with the regular horizon in this linearized setup, and the other is fixed to keep the quark at rest. Note that, in expanding the solution in powers of gradients, it is convenient to keep the full Fourier amplitude of the medium velocity as a multiple. In this way, one can combine it with powers of $\o$ and $\vec k$ to form the hydrodynamic gradients in coordinate space.

The first order gradient corrections to the drag force were studied in \cite{Lekaveckas:2013lha, Rajagopal:2015roa}. At this order $a(\o,k^2,r)=-i\o r^3$ \cite{Bu:2014ena} while $b(\o,k^2,r)$ can be set to zero since it appears in $j_i$ as being multiplied by the second order structure $\vec k \left(k\cdot\b\right)$. Expanding about the zeroth order solution we find
\bea
\pa_\s\vec{x}^{(1,1)}_s=-i\o\vec\b\frac{(1+\s)\left(\tan^{-1}(\s)-\frac{\p}{2}\right)-1}{(1+\s)(1+\s^2)}\, ,
\eea
which, upon setting the quark velocity to zero, agrees with the results of \cite{Lekaveckas:2013lha}. The gradient correction to the time-dependent part of the string profile, considered separately in \cite{Lekaveckas:2013lha}, is now included in the zeroth order profile.

Now the two special non-homogeneous terms on the r.h.s of (\ref{stringeoms}) are canceled and we can write the general form of the higher order gradient corrections in the following simple manner:
\bea
\vec{x}^{(1,n)}_s&=&-2i\o\int_\infty^\s\frac{d\s'}{1-\s'^4}\int_1^{\s'}\,d\s''\,\s''\partial_{\s''}\left(\s''\vec{x}^{(1,n-1)}_s(\s'')\right)+\notag\\
&+&\vec\b\int_\infty^\s\,d\s'\,\frac{a^{(n)}(\s')-\s'^2 a^{(n)}(1)}{\s'^2(1-\s'^4)}-\vec k \left(k\cdot\b\right)\int_\infty^\s\,d\s'\,\frac{b^{(n)}(\s')-\s'^2b^{(n)}(1)}{\s'^2(1-\s'^4)}\,.
\eea
Using this relation one can readily find the second and third order gradient corrections to the string profile. The corresponding expressions are rather lengthy and we present here only their large $\s$ expansions:
\bea
\vec{x}^{(1,2)}_s&\simeq&-\frac{1}{72\s^3}\left[\vec k \left(k\cdot\b\right)+3\vec\b\left(k^2+2\o^2(4-\p+2\log 2\right)\right]+\mathcal{O}\left(\frac{1}{\s^4}\right)\notag\\
\vec{x}^{(1,3)}_s&\simeq&-\frac{i\o^3\vec\b}{72\s^3}\left(\p^2-6(\p-2)\log 2\right)+\notag\\
&+&\frac{i\o\left[\vec k \left(k\cdot\b\right)+3\vec\b\,k^2\right]}{288\s^3}\left(3\p-20+6\log 2\right)+\mathcal{O}\left(\frac{1}{\s^4}\right)\,.
\eea
Now we may substitute the string profile up to third order in the gradient expansion into (\ref{eq:dragforcedef}) to obtain the drag force acting on a heavy quark at rest
\bea
&&-\frac{2\p}{\sqrt{\lambda}}\vec{f}^{~(0+1)}=\left(1+i\o\right)\vec\b\notag\\
&&-\frac{2\p}{\sqrt{\lambda}}\vec{f}^{~(2)}=\frac{1}{8}\left(2\left(\p-4-2\log 2\right)\o^2-k^2\right)\vec\b-\frac{3}{8}\vec k \left(k\cdot\b\right)\notag\\
&&-\frac{2\p}{\sqrt{\lambda}}\vec{f}^{~(3)}=-\frac{i\o^3\vec\b}{24}\left(\p^2-6(\p-2)\log 2\right)+\frac{i\o\left[\vec k \left(k\cdot\b\right)+3\vec\b\,k^2\right]}{96}\left(3\p-20+6\log 2\right)\,,~~~~~~~~~~~~~
\label{f3}
\eea
where the first line, giving the drag force up to first order in gradients, agrees with the results of \cite{Lekaveckas:2013lha} in the zero quark velocity limit. The second and third order corrections are new and, to the best of our knowledge, have never been studied in the literature. In the next section we compare these leading terms in gradient expansion with the approximate solution for the drag force as a function of $\o$ and $\vec k$.

If the quark is forced to move along some trajectory $\vec{x}^{(1)}_{b}(\t,\s)|_{\s\to\infty}=\vec{y}(\t)$ in a static and uniform plasma, the drag force gains additional contributions. Since we are studying a linearized problem, the effects of the medium motion can be separated and one only has to find the homogeneous solution of (\ref{stringeoms}) for the new boundary conditions. To this end, it is convenient to separate the boundary trajectory from the bulk string profile, defining $\vec{x}^{(1)}_{eff}(\t,\s)=\vec{x}^{(1)}_{b}(\t,\s)-\vec{y}(\t)$, and solving the new equations of motion
\bea
\mathcal{D}_{\s}\vec{x}^{(1)}_{eff}=-2i\o\s \vec y~~~,~~~\vec{x}^{(1)}_{eff}(\t,\s)|_{\s\to\infty}=0\,,
\label{stringeomshom}
\eea
where $\vec{y}$ is the Fourier amplitude of the trajectory. Now, one may notice that the string profile for a heavy quark moving along some trajectory (\ref{stringeomshom}) can be mapped to the profile in the medium undergoing the inverse motion, so the quark is effectively moving along the same trajectory with respect to the medium. Indeed, in this case $\vec k=0$, the vector sector of the bulk metric is defined by the exact solution $a=-i\o \s^3$, and (\ref{stringeoms}) reads
\bea
\mathcal{D}_{\s}\vec x^{(1)}_s=-2\s\vec\b~~~,~~~\vec x^{(1)}_s(\t,\s)|_{\s\to\infty}=0\,.
\eea
Thus, the effects of the quark trajectory on the drag force, to linear order in the displacement amplitude, can be obtained directly from the corresponding medium flow. As a result, in what follows we focus on the quark at rest.

\subsection{Matching procedure for the linearized string equations}\label{scheme}
In this section we describe our method for solving the equations of motion for the string profile (\ref{stringeoms}). We adopt the same style of matching procedure as first developed in \cite{Gregory:2009fj}, and then used in \cite{Bu:2014ena}, to provide an approximate semi-analytic scheme of solving the set of ODEs for the $(a,b,c,d)$ coefficient functions -- equivalent to the dynamical Einstein equations. This allows one to determine the viscosities in linearized all order hydrodynamics. We solve for the string profile using the same recursive (and completely algebraic) procedure to determine the coefficients in its power series.

The first step is to define new coordinates that allow for power series expansions about the horizon as well as the boundary. For this we follow \cite{Bu:2014ena} and introduce $u \equiv 1/r$, which maps the domain of our problem from $r \in [1,\infty]$, with the horizon at $r=1$ and the boundary at $r\rightarrow \infty$, to $u \in [0,1]$, with the horizon at $u=1$ but the boundary now at $u=0$. 

We now consider the profile $\vec{x}$ as a function of the new coordinate $u$ and, for convenience, define new rescaled versions of $a$ and $b$ according to 
\begin{eqnarray}
\tilde{a}(\o,k^2,u)=u^4a(\o,k^2,u)~~~,~~~\tilde{b}(\o,k^2,u)=u^4b(\o,k^2,u)\label{ab}\, ,
\end{eqnarray}
as these definitions make the corresponding boundary series regular. Upon such redefinitions, the string equations of motion (\ref{stringeoms}) take the form
\bea
\mathcal{D}_u\vec{x}=\vec{\b}(i\o u-2-u\partial _u\tilde{a}+2\tilde{a})+\vec k \left(k\cdot \b\right)(u\partial _u\tilde{b}-2\tilde{b})\, ,
\label{ustring}
\eea
where $\mathcal{D}_u$ is related to $\mathcal{D}_{\s}$ through the aforementioned coordinate transformation and is defined by
\bea
\mathcal{D}_u\equiv (u^5-u)\partial _u ^2+2(1-i\o u+u^4)\partial _u+2i\o\, .
\eea
The form of (\ref{ustring}) motivates us to parametrize the string profile with the two independent contributions proportional to $\b_i$ and $\pa_i \left(\pa\cdot \b\right)$
\bea
\vec{x}=x_A\, \vec\b-x_B\, \vec k \left(k\cdot \b\right)\label{stringprof} \, ,
\eea
where we choose the subscripts ``$A$'' and ``$B$'' to highlight the fact that this is analogous to the decomposition of the vector sector (\ref{vectorsector}). Furthermore, the form of (\ref{ustring}) tells us that the component $x_A$ communicates only with $\tilde{a}$, and $x_B$ only with $\tilde{b}$. Thus, this decomposition breaks (\ref{ustring}) up into the two equations
\begin{eqnarray}
\mathcal{D}_ux_A&=&-u\partial _u\tilde{a}+2\tilde{a}+i\o u -2 \label{Xaeqn} \\
\mathcal{D}_ux_B&=&-u\partial _u\tilde{b}+2\tilde{b}\label{Xbeqn} \, .
\end{eqnarray}

Next, we expand each sector of the string profile indexed by $I \in \{A,B\}$ in power series about both the boundary at $u=0$ and the horizon at $u=1$, taking the following forms:
\bea
x_I=\sum_{n=0}^{\infty}x^{\rm b} _{I,n}u^n  \, \, , \, \, x_I=\sum_{n=0}^{\infty}x^{\rm h}_{I,n}(1-u)^n\label{seriesexp} \, ,
\eea
where the superscripts ``b'' and ``h'' denote the boundary and horizon expansion coefficients, respectively.

The idea behind the matching method is to generate two power series solutions that solve the equations of motion for the string profile in the bulk geometry -- one in the neighborhood of the horizon, the other in that of the boundary. These neighborhoods do not need to extend over the whole domain, but must have a nonzero overlap. Lifting the restraint of each series providing a solution over the entirety of the domain allows each series to be truncated at some fixed order, while requiring a finite overlap between the two domains of validity allows for the undetermined coefficients to be matched. The process of matching amounts to demanding that the two series, as well as their first derivatives, agree at an arbitrary point of overlap, which, following \cite{Bu:2014ena}, we choose to be $u=1/2$. In our calculations, we find that we are able to construct highly accurate solutions by truncating each series to tenth order. The matching conditions then read:
\begin{eqnarray}
\sum_{n=0}^{10}x^{\rm b} _{I,n}u^n\biggr|_{u=1/2}&=&\sum_{n=0}^{10}x^{\rm h} _{I,n}(1-u)^n\biggr|_{u=1/2} \label{match0} \\
\frac{d}{du}\left[\sum_{n=0}^{10}x^{\rm b}_{I,n}u^n\right]\biggr|_{u=1/2}&=&\frac{d}{du}\left[\sum_{n=0}^{10}x^{\rm h}_{I,n}(1-u)^n\right]\biggr|_{u=1/2}\label{match1} \, .
\end{eqnarray}
Therefore, solving for the string profile amounts to plugging the series expansions (\ref{seriesexp}) into (\ref{Xaeqn}) and (\ref{Xbeqn}), then demanding that the equations are solved at each order. This determines the expansion coefficients $x^{\rm b}_{I,n}$ and $x^{\rm h}_{I,n}$. Each string profile function $x_I$ satisfies a second order equation. There are in general two arbitrary constants defining a solution to be fixed by the boundary conditions. The two boundary conditions are given at the different boundaries of the problem -- the string profile is required to be regular at the worldsheet horizon $u=1$ and stay at rest for $u=0$. Thus, in the boundary expansions we set $x^{\rm b}_{I,0}=0$ and all higher order coefficients can be expressed through the bulk metric parameters and $x^{\rm b}_{I,3}$. Similarly, in the near-horizon expansion, one free parameter is fixed by requiring regularity and all coefficients can be expressed through $x^{\rm h}_{I,0}$. These two free coefficients are then fixed by the matching procedure, where $x^{\rm b}_{I,3}$ is related to the drag force, as we will see below.

Note that in solving for the string profile, the $\tilde{a}$ and $\tilde{b}$ coefficient functions appear in (\ref{Xaeqn}) and (\ref{Xbeqn}), respectively. To solve for these, we employ the exact same scheme as described above, except as applied to the Einstein equations that result from the decomposition of the metic components according to (\ref{vectorsector}). These equations reduce to a set of four coupled ODEs for $(\tilde{a},\tilde{b},\tilde{c},\tilde{d})$, where $\tilde{c}$ and $\tilde{d}$ are the usual $c$ and $d$ functions, just in the coordinate $u$, while $\tilde{a}$ and $\tilde{b}$ are as defined in (\ref{ab}). Since the ODEs are coupled, we perform the aforementioned procedure simultaneously for $(\tilde{a},\tilde{b},\tilde{c},\tilde{d})$. That is, we utilize near-boundary and near-horizon expansions for each of the four coefficient functions truncated at tenth order, just as done for the string profile. Our results for the metric functions agree with the results of \cite{Bu:2014ena} and with the small momenta expansion.

With these expansions at hand, the definition (\ref{eq:dragforcedef}) enables us to directly relate the drag force to the boundary coefficients of $\vec{x}$, $\tilde{a}$, and $\tilde{b}$. To do so, we decompose the drag force in the same manner as the string profile (\ref{stringprof}) 
\bea
-\frac{2\p}{\sqrt{\l}}\vec{f}= f_A\,\vec  \b-f_B\, \vec k \left(k\cdot \b\right)  \label{forcedecomp}\, ,
\eea
and find that the components $f_A$ and $f_B$ have the following relations to the boundary expansion coefficients:
\bea
f_A&=&3\,x^{\rm b}_{A,3} - a^{\rm b}_2 \notag \\
f_B&=&3\,x^{\rm b} _{B,3} - b^{\rm b}_2 \label{fcoeffs}
\eea
where $a^{\rm b}_2=0$ and $b^{\rm b}_2=-1/3$, see \cite{Bu:2014ena}. 

We close this section by reiterating that the matching method described here yields an approximate solution to the string equations of motion (\ref{Xaeqn}) and (\ref{Xbeqn}). In order to assess the stability of such solutions, we further solve these equations numerically, utilizing well-known spectral methods for solving ODEs \cite{Boyd:2001, NumRecC:2007}. We find that the approximate solutions obtained via the matching procedure indeed agree well with the numerical solutions obtained through spectral methods, see Figs. \ref{appendixplot1} and \ref{appendixplot2} in the appendix.

\subsection{Drag force in all order linearized hydrodynamics}
\begin{figure}[t!]
\begin{center}
\includegraphics[width=2.95in]{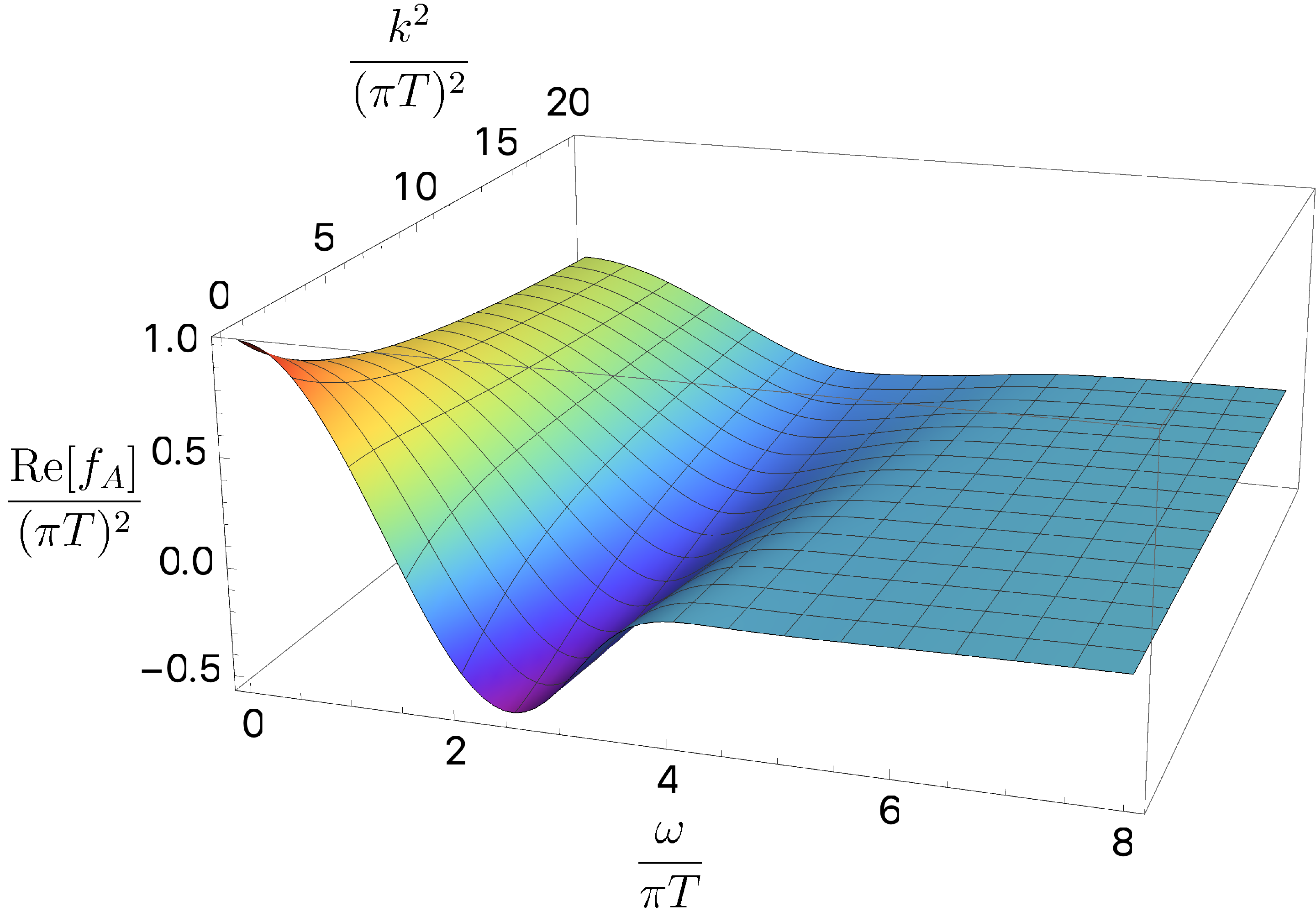}
\includegraphics[width=2.95in]{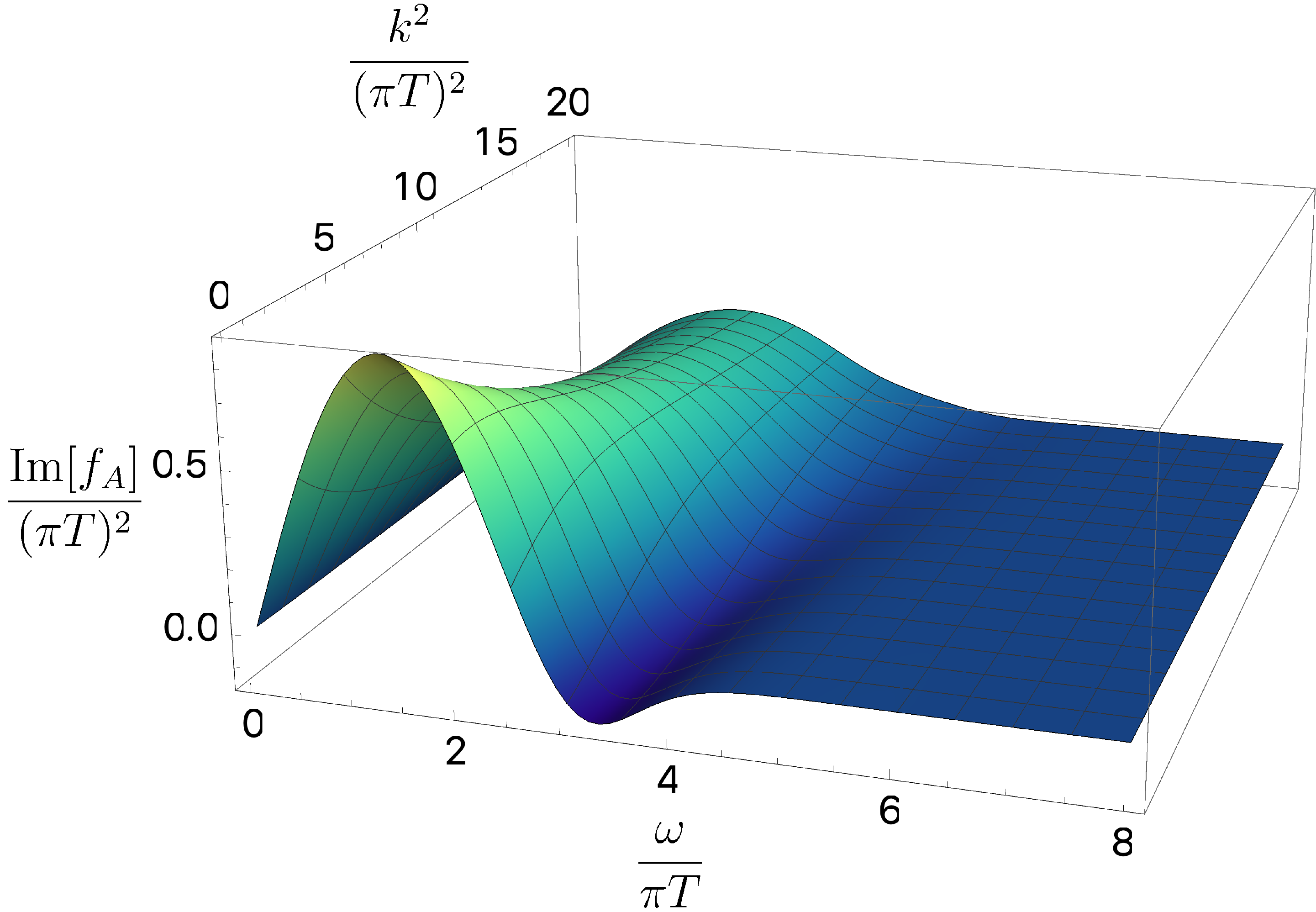}
\includegraphics[width=2.95in]{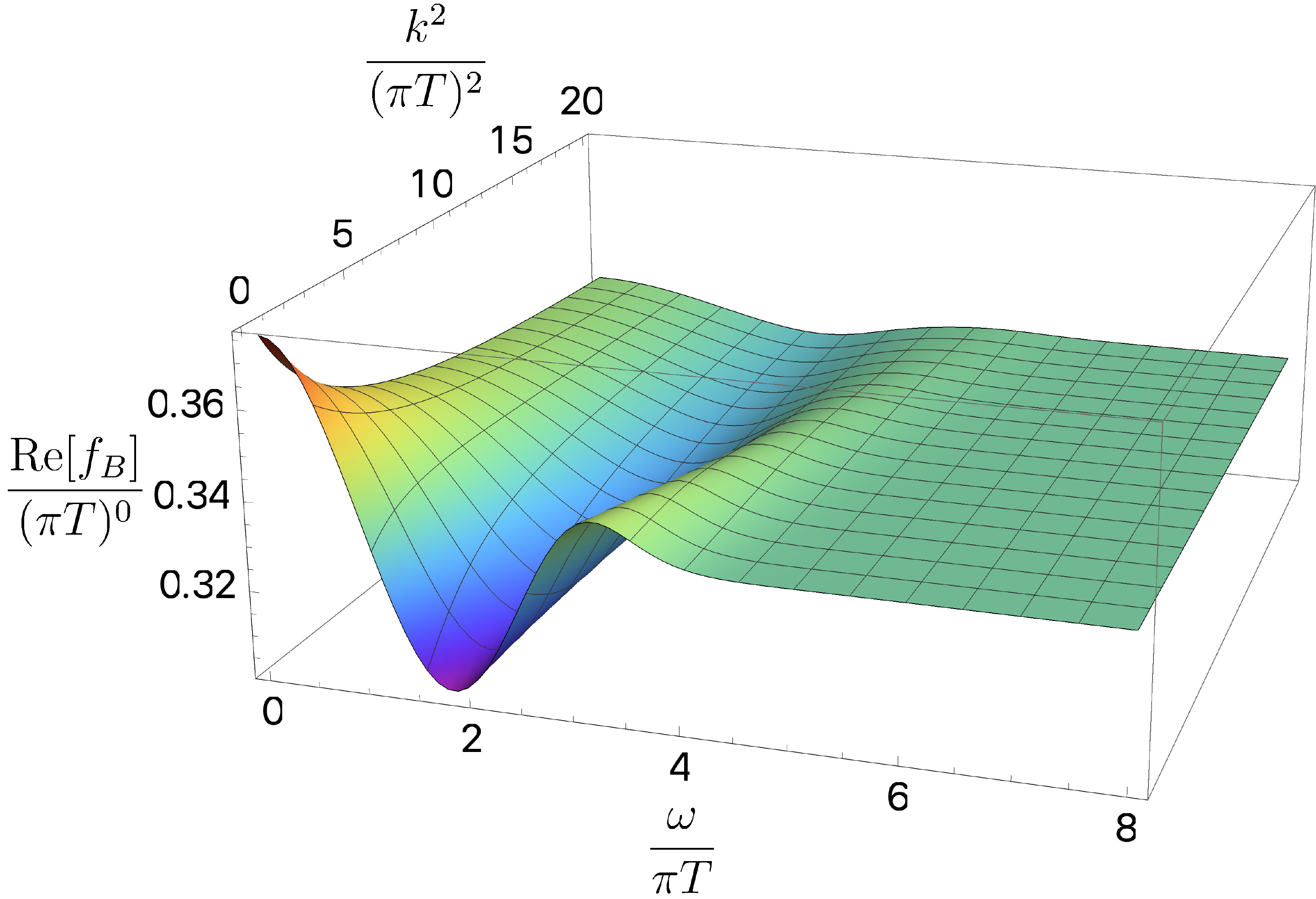}
\includegraphics[width=2.95in]{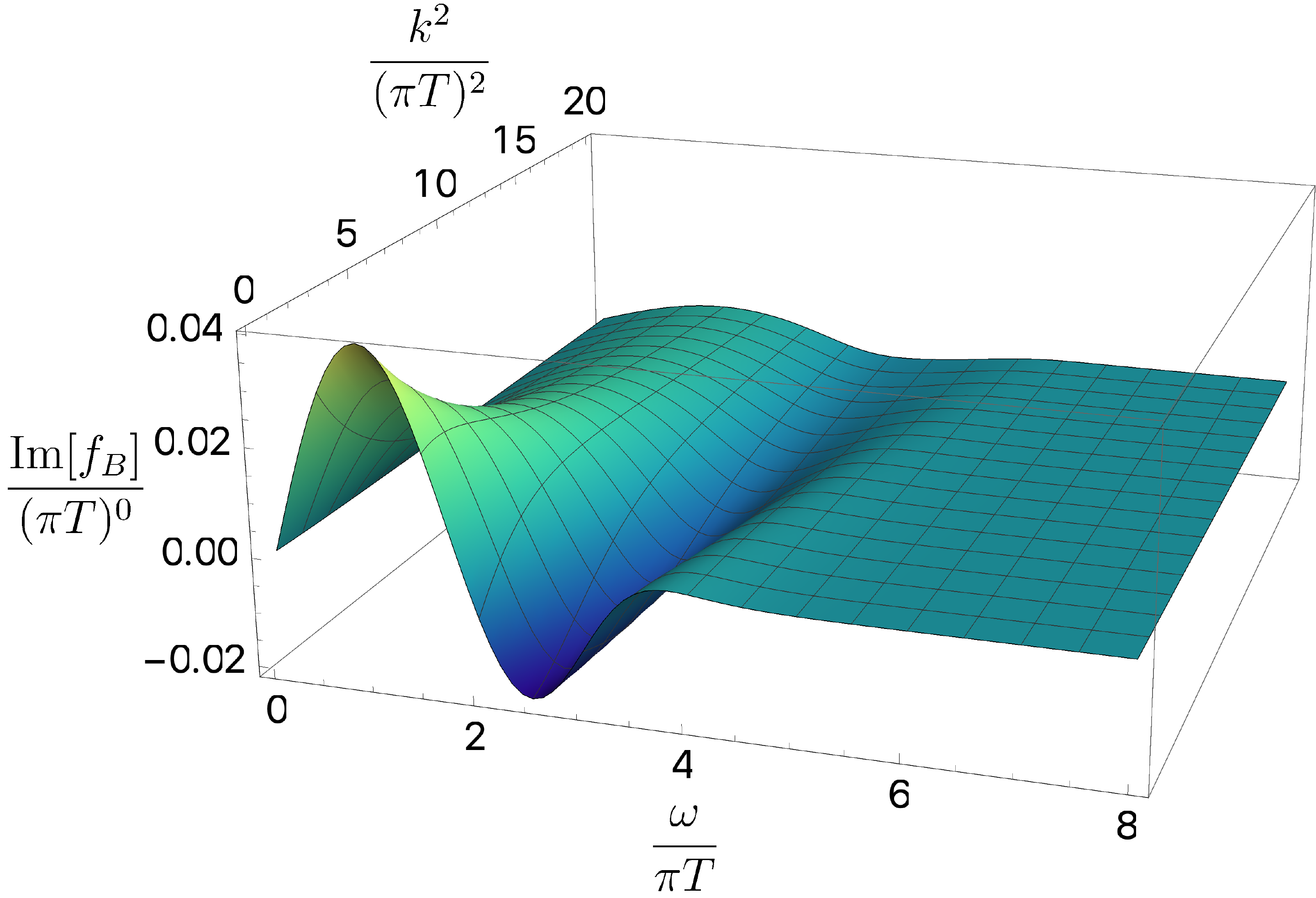}
\caption{Real and imaginary parts for the $A$ component (above) and $B$ component (below) of the all order drag force as functions of $\o$ and $k^2$. Note that each quantity is displayed in units of $\p T$ based on their dimensionality.}
\label{ReImf}
\end{center}
\end{figure}

In this section we present our results obtained for the all order drag force as functions of $\o$ and $k^2$, utilizing the scheme discussed in Section \ref{scheme}, which are plotted in Fig. \ref{ReImf}. For illustrative purposes, we re-introduce powers of $\p T$, which provide the proper dimensionalities for each physical unit. Note that, since $[\w],[k_i]\sim\p T$ and $[f_i]\sim (\p T)^2$, the decomposition (\ref{forcedecomp}) implies that $[f_A]\sim (\p T)^2$ and $[f_B]\sim (\p T)^0$.

One may immediately notice that most of the functions in Fig. \ref{ReImf} vanish as $\w$ and $k^2$ become large, as expected for viscosities, see \cite{Bu:2014ena}. This is intuitive, since, as $\w$ increases, the rate at which the fluid fluctuates about the impurity will eventually become so large as to prevent the impurity from being able to resolve any individual disturbance. Strikingly, the real part of $f_B$ is the only component which does not approach zero with increasing $\w$ and $k^2$, but rather reaches a constant value $1/3$, which comes directly from $b_2^{\rm b}$ in (\ref{fcoeffs}). This constant drag force, however, can be removed in another frame moving relatively to the Landau frame where the quark feels no drag force at rest. Indeed, while $f_A$ tends to zero at large frequencies, it still can compensate the effect of a constant $f_B$ if we assume the quark to be moving with a velocity amplitude growing at large $\o$. In the new frame, such that the quark undergoing the terminal motion in the Landau frame is at rest, one, in general, expects to see a non-zero energy-momentum transfer, which results in no drag on the impurity\footnote{This situation is similar to the energy-momentum transfer caused by the chiral effects, see \cite{Rajagopal:2015roa}. In that case the drag force is also non-zero in the Landau frame and the heavy quark is forced to move with a constant terminal velocity. Upon boosting to the frame moving with this terminal velocity one finds that the quark feels no drag despite the charge and energy-momentum currents flowing in this no-drag frame.}. In this sense, the energy-momentum transfer is non-dissipative \cite{Rajagopal:2015roa}.  

It should be also mentioned that the drag coefficients are expected to vanish in the limit of large $\w$ if derived in the microscopic theory, which is usually defined in the rest frame of the entropy density, c.f. \cite{Rajagopal:2015roa}. An explicit calculation shows that the entropy current, which is related to the $j_i$ component of the bulk metric evaluated at the horizon \cite{Chapman:2012my,Eling:2012xa}, has a non-vanishing contribution proportional to $\pa_i(\pa\cdot\b)$ at large $\w$ in the Landau frame. However, the behavior of the drag force in the general frame of linearized all order hydrodynamics requires further investigation.

\begin{figure}[t!]
\begin{center}
\includegraphics[width=2.95in]{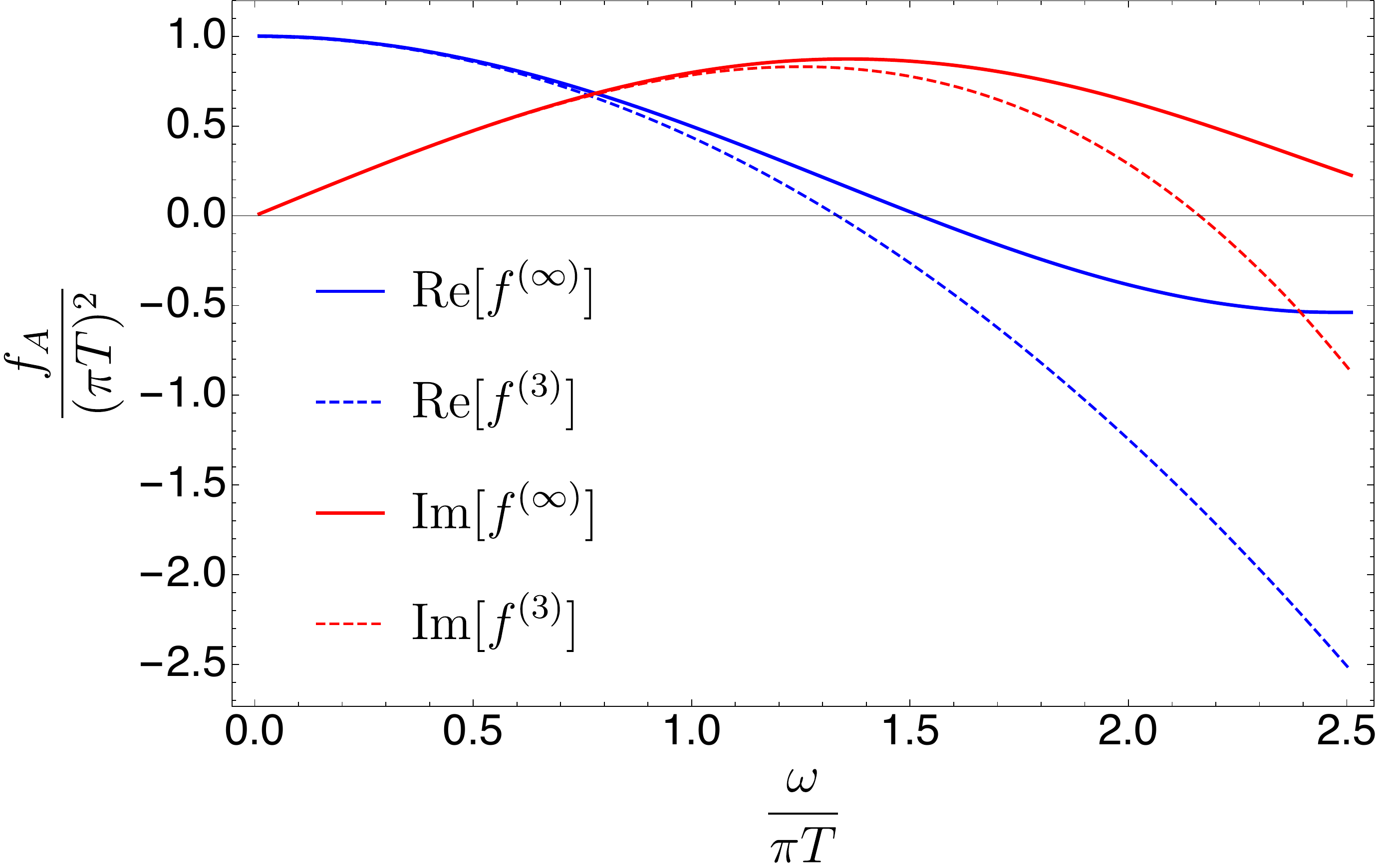}
\includegraphics[width=2.95in]{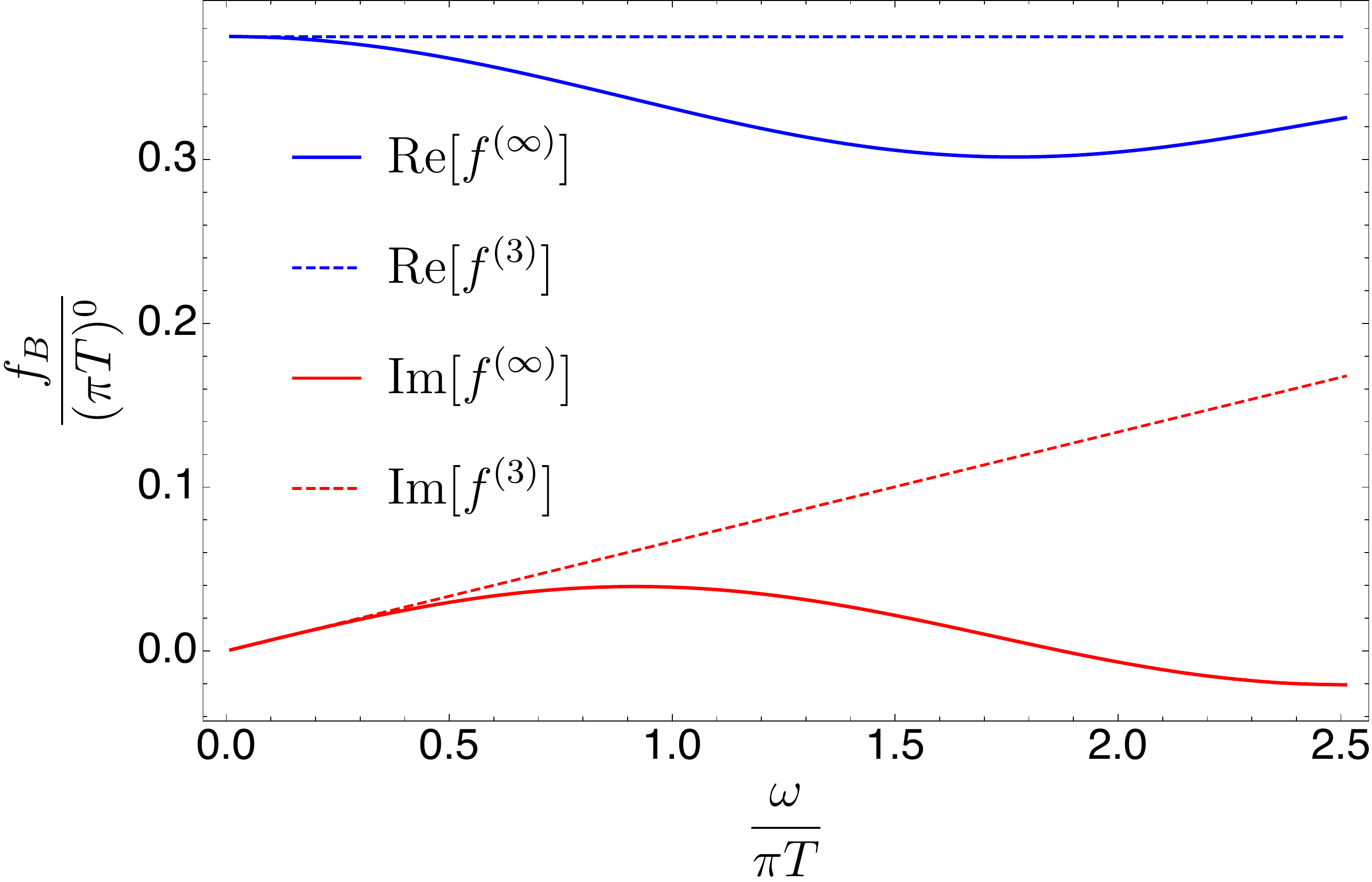}
\caption{Real and imaginary parts for the all order as well as third order drag force for the $A$ component (left) and the $B$ component (right) at fixed $k^2=0$ as functions of $\o$. The all order drag force is depicted by solid lines labeled $f^{(\infty)}$, while the dotted lines labeled $f^{(3)}$ represent the drag force obtained to third order in the gradient expansion (\ref{f3}).}
\label{f3AABB}
\end{center}
\end{figure}
In the hydrodynamic limit of small $\w$ and $k^2$, the approximate all order results can be compared to the analytic expressions for the drag force (\ref{f3}) derived up to the third order. Note that $f_B$ contributes starting at second order, since it is scaled by $\pa_i(\pa\cdot\b)$ in (\ref{forcedecomp}) -- it can thus be ignored in the limit $\w,k^2\rightarrow0$. In this limit, $f_A\rightarrow1$ with identically zero imaginary part, meaning that the drag force felt by the static quark is along the direction of the fluid velocity, exactly as it should be. The case of small, but finite, gradients is shown in Fig. \ref{f3AABB}, where, for the purposes of clarity, we examine the $\w$ dependence of the drag force in the $k^2=0$ slice. The two plots depict the ranges of convergence for the fixed and all order calculations. For the case of $f_A$, this range is remarkably wide -- a good agreement exists all the way up to $\w \sim \p T$. This is noteworthy as the gradient expansion for hydrodynamics is typically expected to be valid in the regime of $\w \ll T$. For $f_B$, the only contribution that is exactly third order must be obtained via a factor of $\w$, which then inevitably includes a factor of $i$. Hence, to third order in gradients, the real part of $f_B$ is expected to be constant while the imaginary part varies linearly with $\w$ -- exactly as observed in Fig. \ref{f3AABB}. Another illustration can be obtained by expanding the approximate analytic expressions for the all order drag force in the double limit of small $\w$ and $k^2$ to third order
\bea
f_A(\w,k^2)&=&1.00  +1.00 \,i \w -0.126 \,k^2 -0.561 \,\w^2  -0.202 \,i\w k^2  -0.213\,i \w^3 +\dots \notag \\
 f_B(\w,k^2)&=& 0.375 +0.0674 \,i \w +\dots  \, ,
\eea
which are in agreement with the results obtained via the gradient expansion (\ref{f3}).

The all order formalism allows one to obtain the behavior of the drag force in the limit of large $\w$, similar to the case of viscosities \cite{Bu:2014ena}, providing information that is unattainable using the traditional gradient expansion. By performing this expansion on each component of the all order drag force we obtain:
\bea
f_A(\w,k^2)\simeq-\frac{1}{\w^2}~~~,~~~f_B(\w,k^2)\simeq \frac{1}{3}-\frac{10^{-1}}{\w^4}
\eea
as $\w\rightarrow\infty$.
 
\section{Summary and outlook}
\label{sec:Summary}
In this paper we have studied the drag force exerted on a heavy quark moving through an evolving strongly coupled plasma, motivated by the physics of the QGP produced in HIC. The QGP is similar to the $\mathcal{N}=4$ SYM plasma in the strongly coupled regime and we have used the latter as a model. We have described the $\mathcal{N}=4$ SYM plasma in the large-$N_c$ limit and at strong coupling within the AdS/CFT correspondence, which allows it to be related to the dual higher dimensional classical gravity. In the bulk theory, a heavy quark corresponds to the end point of a classical string which, in the infinite mass limit, is attached to the AdS boundary and hangs down to the horizon of a black brane \cite{Herzog:2006gh, Gubser:2006bz}, which is needed to introduce a finite temperature. By studying this string profile in the background bulk metric that corresponds to the flowing strongly coupled plasma, we have derived the drag force.

Previously, medium motion effects on heavy quark energy loss were considered up to first order in the hydrodynamic gradient expansion \cite{Lekaveckas:2013lha, Rajagopal:2015roa}. We have extended these previous studies of the drag force in an evolving plasma to include higher order gradient effects and, in fact, re-summed the gradient contributions that are linear in the medium velocity to all orders using the  linearized fluid/gravity correspondence \cite{Bu:2014ena}. The equations of motion for the string profile in this linearized background metric can be solved analytically order by order in gradients if such an additional expansion is performed. Using this procedure, we have solved for the string profile and the drag force exerted on the heavy quark up to the third order in gradients (\ref{f3}). Our results agree with the drag force obtained in \cite{Lekaveckas:2013lha, Rajagopal:2015roa} up to the first order in gradients, upon setting the quark velocity to zero. We have further used a matching procedure, that relates the series expansions of the bulk metric and string profile at the boundary and horizon, to derive approximate solutions incorporating gradient effects in the linearized fluid/gravity correspondence to all orders, see Fig.\,\ref{ReImf}. Thus, our study provides further insight into how both hydrodynamic effects and more general medium perturbations modify the energy loss of a heavy quark in strongly coupled plasmas. The approximate all order results also allow one to study how the drag force on a heavy quark behaves in the limit of rapidly varying linear perturbations, which is opposite to the hydrodynamic limit. 

In \cite{Bu:2014ena}, it is argued that the dissipative transport coefficients $\eta(\o, k^2)$ and $\zeta(\o, k^2)$, both vanish in the limit of large $\w$ and $k^2$ since there is no response at very short time/length scales. The same argument applies to the drag force, but special care must be taken in interpreting the limiting behavior of the $f_B$ component. Curiously, we have found that the drag force contribution $f_B$, proportional to $\pa_i(\pa\cdot\b)$, in fact goes to a constant value as $\w$ increases. This indicates that a heavy quark is forced to move in the Landau frame and the drag force goes to zero at large $\w$ in the frame corresponding to that motion. What is less obvious is that, in this frame and in the same limit, there is in general a non-zero energy-momentum transfer which is, in fact, non-dissipative -- an impurity placed in its flow feels no drag, for a discussion see \cite{Rajagopal:2015roa}. In the more general case, where conductivities due to external forces are considered, there are other examples of kinetic coefficients not vanishing at large frequencies, see e.g. \cite{Bu:2015ika}. It will be interesting to study the frame dependence of the large $\o$ behavior for the all order drag force and these kinetic coefficients in a unified way. We leave this to future studies. 

\section{Acknowledgements}
The authors would like to thank K. Bazarov, who participated at early stages of this study. The authors are grateful to J. Brewer, Z. B. Kang, M. Lublinsky, K. Rajagopal, W. van der Schee and I. Vitev for fruitful discussions. J. R. would like to thank the Theoretical Division of Los Alamos National Laboratory for its hospitality during the completion of this work. The work of J. R. is supported by the UC Office of the President through the UC Laboratory Fees Research Program under Grant No. LGF-19-601097. The work of A. S. is partially supported through the LANL/LDRD Program. A.S. is also grateful for support by RFBR Grant 18-02-40056 at the beginning of this project.

\appendix
\renewcommand{\thefigure}{\thesection.\arabic{figure}}
\renewcommand{\theHfigure}{\thesection.\arabic{figure}}
\setcounter{figure}{0}
\section{Fully numerical solutions}

In this appendix we provide our results obtained for the all order drag force as functions of $\o$ and $k^2$, utilizing the fully numerical methods mentioned at the end of Section \ref{scheme}. 

In order to numerically solve the string equations of motion, (\ref{Xaeqn}) and (\ref{Xbeqn}), we first numerically solve the equations of motion for the metric functions $(a,b,c,d)$, following the same methods as described in \cite{Bu:2014ena}. Namely, the ODEs are solved using traditional spectral methods \cite{Boyd:2001, NumRecC:2007}. To fulfill the boundary conditions for the string profile, we employ the same shooting procedure -- trial solutions for the string profile with varying finite values at the $u=1-\epsilon$ stretched worldsheet horizon are tuned until they meet the desired near-boundary values at $u=\epsilon$, where $\epsilon>0$ is introduced to avoid numerical instabilities. This allows us to meet the boundary conditions to a high numerical accuracy ($\sim 10^{-14}$). We refer the reader to \cite{Bu:2014ena} for a detailed description of the shooting method as applied to the spacetime geometry at hand.

\begin{figure}[h]
\begin{center}
\includegraphics[width=2.95in]{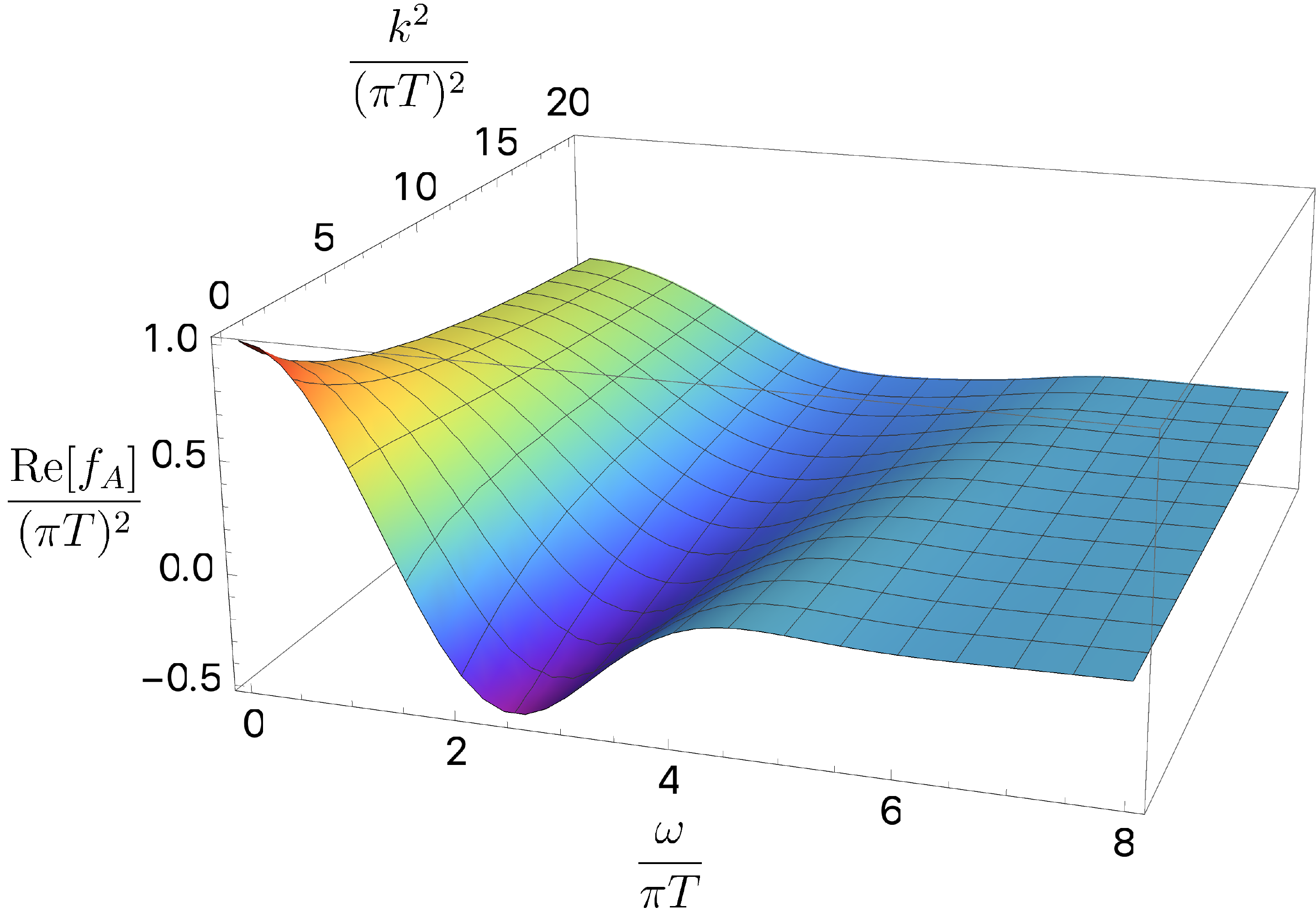}
\includegraphics[width=2.95in]{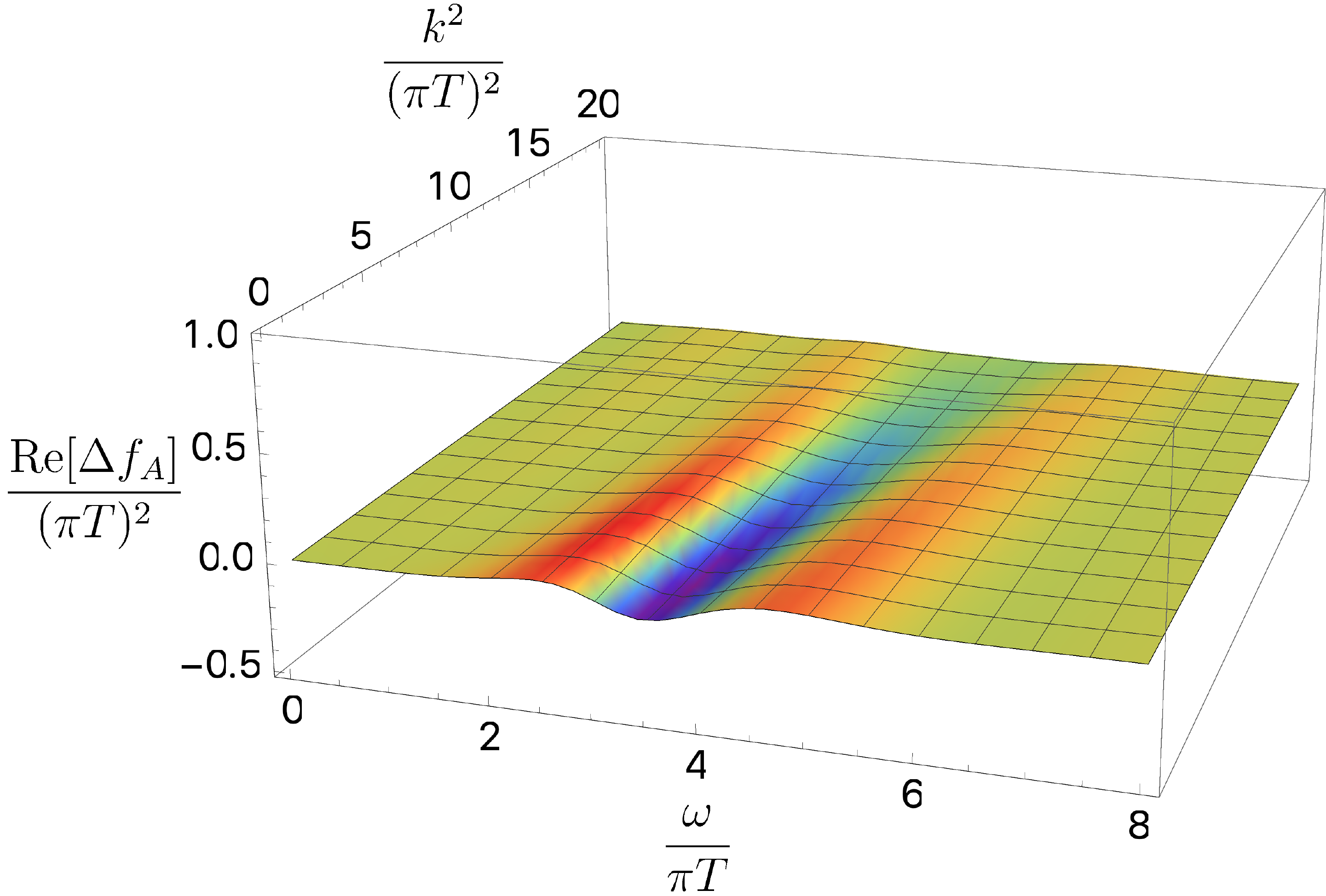}
\includegraphics[width=2.95in]{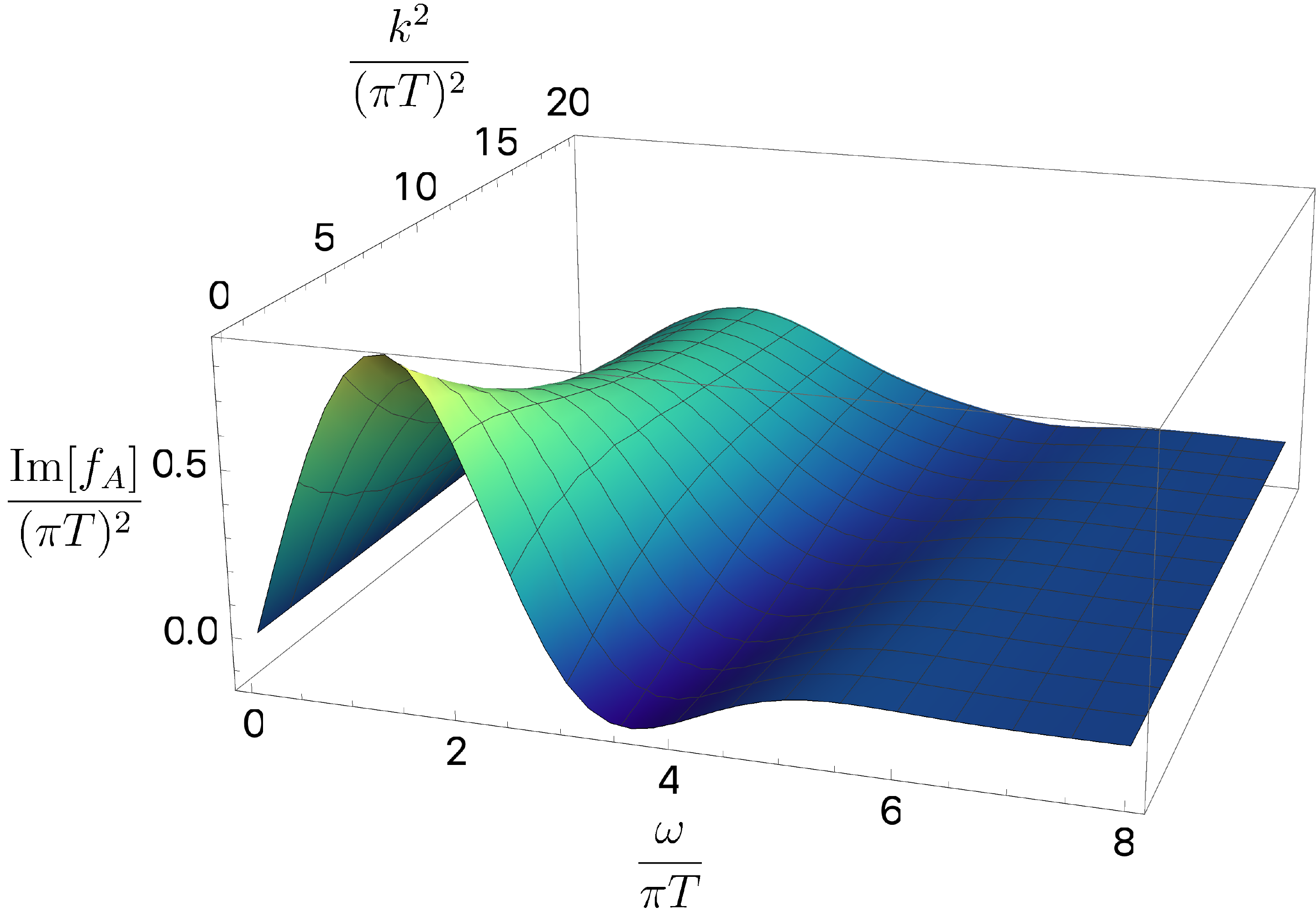} 
\includegraphics[width=2.95in]{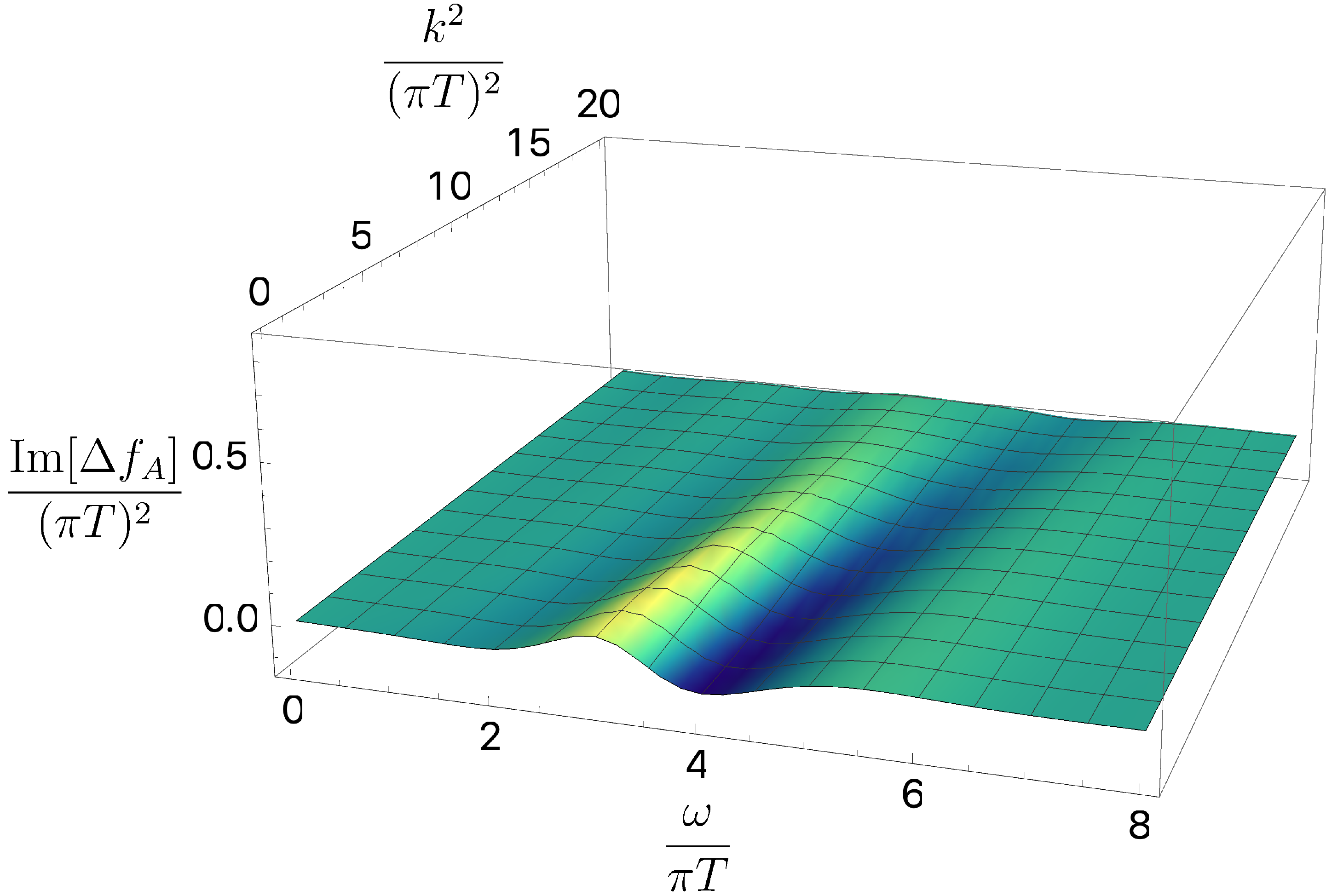}
\caption{Left column: Real and imaginary parts for the $A$ component of the all order drag force as functions of $\o$ and $k^2$, evaluated with the spectral method. Right column: The resulting differences between the $A$ components as evaluated through the spectral and matching methods.}
\label{appendixplot1}
\end{center}
\end{figure}

\begin{figure}[t]
\begin{center}
\includegraphics[width=2.95in]{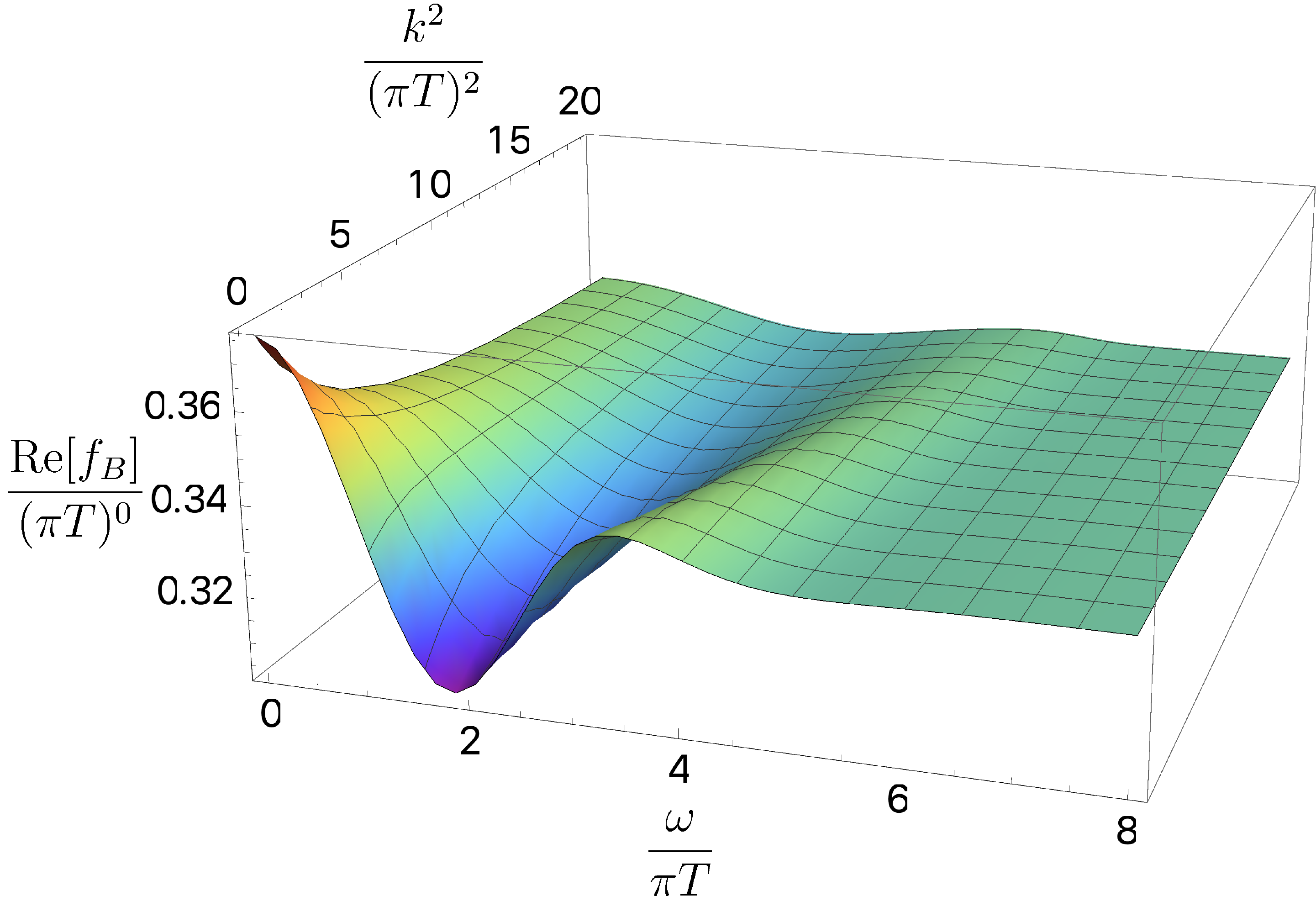}
\includegraphics[width=2.95in]{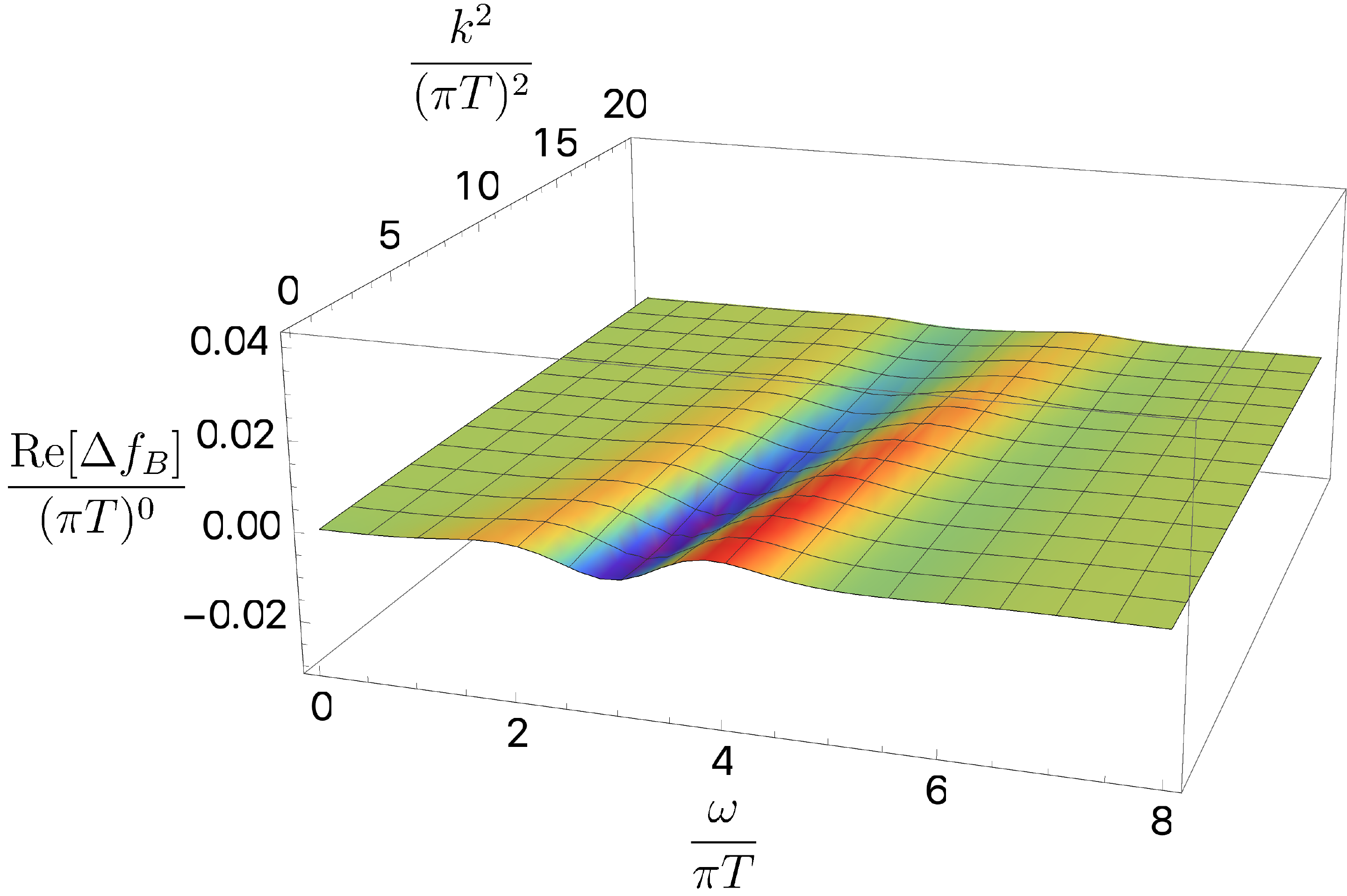}
\includegraphics[width=2.95in]{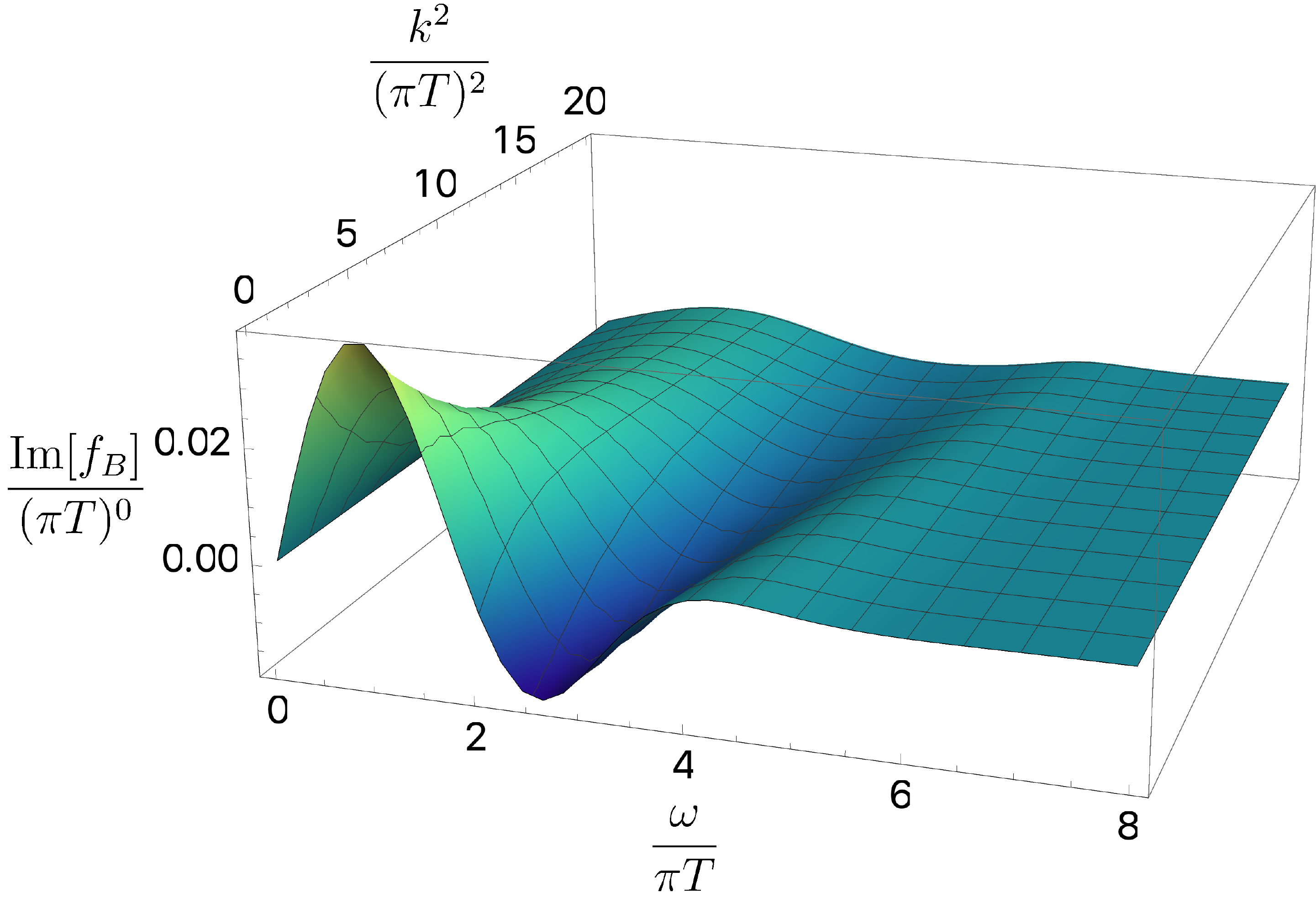} 
\includegraphics[width=2.95in]{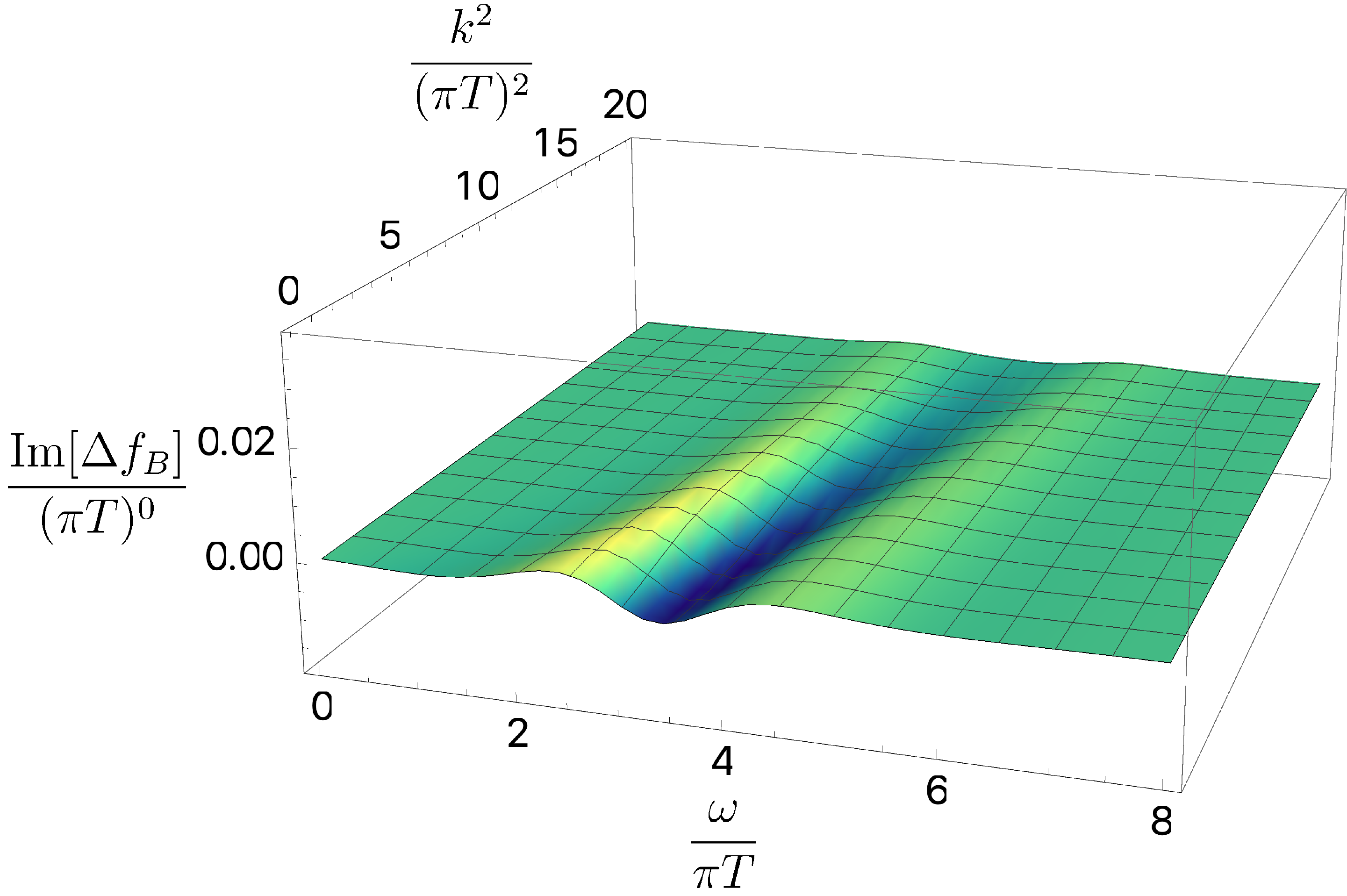}
\caption{Left column: Real and imaginary parts for the $B$ component of the all order drag force as functions of $\o$ and $k^2$, evaluated with the spectral method. Right column: The resulting differences between the $B$ components as evaluated through the spectral and matching methods.}
\label{appendixplot2}
\end{center}
\end{figure}

We display the numerical solutions in the left columns of Figs. \ref{appendixplot1} and \ref{appendixplot2}. By comparing these plots to those of Fig. \ref{ReImf}, we see that the approximate solutions displayed in Fig. \ref{ReImf} are quite robust approximations to the numerical solutions -- the discrepancies are only minute shifts localized in small regions of the displayed surfaces. In order to identify the precise locations of these minor discrepancies, the right columns of Figs. \ref{appendixplot1} and \ref{appendixplot2} display the differences that correspond to the neighboring plots in the left columns, defined as $\Delta f_{A,B} = f_{A,B}^{\rm spectral}- f_{A,B}^{\rm matching}$, where $ f_{A,B}^{\rm spectral}$ denotes the drag force solutions obtained through the fully numerical spectral methods and $f_{A,B}^{\rm matching}$ denotes the approximate solutions obtained through the matching procedure of Sec. \ref{scheme}. These differences are attributed to the higher accuracy of the solutions obtained through spectral methods as compared to the matching procedure truncated at tenth order.

\clearpage
\bibliographystyle{bibstyle}
\bibliography{DF}

\end{document}